\documentclass{emulateapj}
\usepackage{natbib}
\usepackage{epsf,graphics}
\usepackage{subfigure,graphicx}
\usepackage{bm}

\begin{document}

\title{Correcting velocity dispersions of dwarf spheroidal galaxies for binary 
orbital motion}

\author{Quinn E. Minor, Greg Martinez, James Bullock, Manoj Kaplinghat}
\affil{Department of Physics and Astronomy, University of California, Irvine CA 
92697, USA}
\author{Ryan Trainor}
\affil{Department of Astronomy, California Institute of Technology, Pasadena CA 
91125, USA}

\begin{abstract}
We show that measured velocity dispersions of dwarf spheroidal galaxies from 
about 4 to 10 km/s are unlikely to be inflated by more than 20\% due to the 
orbital motion of binary stars, and demonstrate that the intrinsic velocity 
dispersions can be determined to within a few percent accuracy using two-epoch 
observations with 1-2 years as the optimal time interval.  The crucial 
observable is the threshold fraction---the fraction of stars that show velocity 
changes larger than a given threshold between measurements. The threshold 
fraction is tightly correlated with the dispersion introduced by binaries, 
independent of the underlying binary fraction and distribution of orbital 
parameters.  We outline a simple procedure to correct the velocity dispersion 
to within a few percent accuracy by using the threshold fraction and provide 
fitting functions for this method.  We also develop a methodology for 
constraining properties of binary populations from both single- and two-epoch 
velocity measurements by including the binary velocity distribution in a 
Bayesian analysis.
\end{abstract}


\section{Introduction}\label{sec:intro}

In recent years, a large number of dwarf spheroidal satellite galaxies of the 
Milky Way have been discovered using the Sloan Digital Sky Survey 
(\citealt{willman2005}, \citealt{zucker2006}).  These galaxies are much fainter 
than previously known Milky Way satellites, having larger mass-to-light ratios 
and velocity dispersions that range from 7.6 km/s down to 3.3 km/s 
(\citealt{simon2007}).  Estimating the amount of dark matter contained in these 
galaxies becomes more susceptible to error than in the larger dwarf 
spheroidals.  This is not only because of the statistical error associated with 
small stellar samples, but also because sources of velocity contamination 
constitute a larger fractional error due to the galaxy's small intrinsic 
velocity dispersion.  Potential sources of contamination come from foreground 
Milky Way stars, atmospheric jitter in red giant stars (\citealt{pryor1988}), 
and an inflated velocity dispersion due to the orbital motion of binary stars 
(\citealt{olszewski1996}).  Among these, binaries have been the most difficult 
to correct for because the binary distribution of velocities in environments 
beyond the solar neighborhood is not well known and difficult to constrain 
without a large number of high-precision radial velocity measurements.

The most prominent signature of binary stars in these galaxies is a 
high-velocity ``tail'' in the velocity distribution due to short-period 
binaries, which gives the distribution a higher kurtosis than that expected 
from a Gaussian. Because of this departure from Gaussianity, the intrinsic 
velocity dispersion is usually estimated by using a robust estimator such as 
the biweight (\citealt{beers1990}), by discarding velocity outlier stars from 
the data sample, or by a combination of both approaches (cf.  
\citealt{mateo1991}).  While these techniques eliminate the largest component 
of the binary dispersion, binaries inflate the first-order (Gaussian) component 
of the dispersion as well.  In previous studies, \cite{olszewski1996} and 
\cite{hargreaves1996} used Monte Carlo simulations to show that the dispersion 
estimated by these techniques is inflated due to binaries by an amount which is 
small compared to the statistical error in their data sets.  Since that time, 
the number of stars in then-known dwarf spheroidals with spectroscopic data has 
increased considerably, from less than 100 at the time of \cite{olszewski1996} 
to more than 1000 for Draco, Fornax, and Carina (\citealt{walker11-09}). For 
such large samples, we will show that the bias in the first-order dispersion 
due to binaries is larger than the statistical error.  More importantly, for 
galaxies whose intrinsic dispersion is small, the first-order dispersion may be 
inflated by somewhat more than the statisical error even in data sets as small 
as 100 stars.

To improve previous estimates of the intrinsic dispersion for these cases, it 
is necessary to model the velocity distribution of the binary population and 
investigate the behavior of the binary dispersion as model parameters such as 
binary fraction are varied.  With this approach it is desirable to find the 
best possible constraints on the binary velocity distribution, and observations 
taken at multiple epochs are very useful for this purpose.  In addition, 
estimating the binary fraction and other binary properties in galaxies and 
clusters beyond the solar neighborhood is useful in its own right, since this 
has been difficult to predict accurately in simulations (\citealt{tohline2002}, 
\citealt{goodwin2006}).  Binaries also affect higher-order moments of the 
velocity distribution, e.g.  kurtosis, which has been shown in principle to 
provide useful information about the mass profile of galaxies 
(\citealt{lokas2005}).

In section \ref{sec:derivations} we derive an analytic formula for the 
center-of-mass velocity distribution of binary stars. In section 
\ref{sec:likelihoods} we use this distribution to derive likelihoods for both 
single- and two-epoch velocity data. We demonstrate in section 
\ref{sec:multi_epoch} how properties of a binary population can be constrained 
from multi-epoch data.  Fig.~\ref{fig:dvchist_dt} shows that repeat 
measurements with a baseline of 1-2 years is sufficient to constrain the binary 
population. We will prove that over a range of velocities of a few km/s, the 
binary fraction is nearly degenerate with the parameters characterizing the 
distribution of periods, and in section \ref{sec:correcting_dispersion} we use 
this fact to develop a model-independent method for correcting the dispersion 
due to binaries by using multi-epoch data.  Fig.~\ref{fig:fdisp_sigreal} 
encapsulates our main result, from which we conclude that the velocity 
dispersions of dwarf spheroidals are unlikely to be inflated by more than 20\% 
by binaries. The procedure to correct the dispersion for binaries is summarized 
at the end of section \ref{sec:correcting_dispersion}.  In section 
\ref{sec:bayesian} we outline a method to combine single- and multi-epoch data 
in a Bayesian analysis, and discuss the issue of foreground contamination and 
how this affects the apparent binary population.  In section 
\ref{sec:fitting_function} we find a fitting function for the velocity 
distribution of binary stars and use it to derive an analytic formula for the 
binary dispersion.  Finally, in section \ref{sec:period_dist} we further 
explore the degeneracy of model parameters and discuss prospects for 
constraining the period distribution of binary systems from multi-epoch 
velocity data.  

\section{Distribution of velocities in the center-of-mass frame of binary 
systems}\label{sec:derivations}

Spectroscopic velocity measurements of binary stars are typically dominated by 
the primary star due to the difference in luminosities and spectra between the 
two stars. This is especially true for stars that lie on the red giant branch, 
for which the luminosity-mass relation steepens drastically. By way of 
comparison, \cite{duquennoy1991} found the ratio of secondary-to-primary masses 
of binary stars within the solar neighborhood to peak at $m/M \approx 0.23$ 
with only $\approx 20\%$ of the stars having mass ratios larger than 0.6.  For 
systems with a mass ratio of 0.6 and the primary lying near the base of the red 
giant branch, the luminosity ratio would be at most $l \approx (m/M)^{2.3} 
\approx 0.3$; this ratio becomes much smaller if the primary star is further 
along on the red giant branch.  Thus, although the cross correlation function 
of the stellar spectra may be double-peaked (\citealt{tonry1979}), the spectrum 
of the primary star will likely dominate the signal unless the mass ratio is 
close to 1, in which case the velocities of the two stars will be nearly equal.  
In view of the preceding arguments, for the remainder of this paper we will use 
the velocity of the primary star in modeling the spectroscopic velocities of 
binary systems.

To model the velocity distribution of the primary star in a population of 
binary systems, first we must find its velocity distribution in the 
center-of-mass frame of the binary.  The motion of two stars orbiting each 
other can be simply expressed in terms of four parameters: the semi-major axis 
$a$, eccentricity $e$, period $P$, and orbital angle $\phi$ with respect to the 
center of mass.  If we also specify the orientation of the orbital plane in 
terms of Euler angles, with the $z$-axis pointing along the line of sight, the 
line-of-sight velocity of the primary star in the center-of-mass frame is given 
by

\begin{equation}
\label{vz_formula}
v_z = \frac{2\pi a}{P}(1-e^2)^{-\frac{1}{2}} g_e(\theta,\psi,\phi),
\end{equation}

\begin{equation}
\label{g_formula}
g_e(\theta,\psi,\phi) \equiv \sin \theta\left[\cos(\psi - \phi) + e \cos(\phi)\right]
\end{equation}
where $a$ is the semi-major axis of the primary star's orbit, and $\theta$ and 
$\psi$ are the second and third Euler angles, respectively. The azimuthal Euler 
angle does not appear explicitly because $v_z$ is invariant under rotations 
about the $z$-axis.

Now taking the log of eq.~\ref{vz_formula} and subsituting Kepler's third law, 
we find

\begin{equation}
\label{P_k_vz}
\log P ~ = ~ k ~-~ 3 \log |v_z|,
\end{equation}

\begin{equation}
\label{k_formula}
k \equiv \log \left\{ 
\left(\frac{|g_e(\theta,\psi,\phi)|}{\sqrt{1-e^2}}\right)^3 \frac{(2\pi q)^3 
m}{(1+q)^2} \right\}
\end{equation}
where $P$ is in years, $m$ is the mass of the primary star in solar masses, $q$ 
is the ratio of secondary-to-primary mass, and $v_z$ is in units of AU/year.  

As an aside, we note that eqs.~\ref{P_k_vz} and \ref{k_formula} can be used to 
make a back-of-the-envelope estimate of the velocity scale associated with a 
given orbital period. By averaging over orientations, mass ratio and 
eccentricity (whose PDF's are given below), we find

\begin{equation}
\label{envelope_estimate}
|v_z| \approx (5.7 ~ \textrm{km/s}) 
\left(\frac{M/M_\odot}{P/\textrm{year}}\right)^{\frac{1}{3}}
\end{equation}
where $M$ is the mass of the primary star. In dwarf spheroidal galaxies and 
globular clusters where red giants have masses $M \approx 0.8 M_\odot$, the 
above estimate shows that periods longer than a few decades will yield 
velocities less than 2 km/s.  Meanwhile, velocities larger than 10 km/s will be 
dominated by binaries with periods shorter than 1 month. The above estimate is 
somewhat sensitive to the mass ratio; whereas we used the approximate mean 
value $q \approx 0.4$ in deriving eq.~\ref{envelope_estimate}, if a system has 
a mass ratio $q \approx 0.8$, the coefficient in front becomes $\approx$ 10 
km/s.

To find a distribution in the center-of-mass velocity $v_z$, the distribution 
of orbital periods must be averaged over all the parameters in 
eq.~\ref{k_formula}.  We must also average over the time taken to traverse one 
orbital cycle, with all times being weighted equally. We express the orbital 
angle $\phi$ in terms of the eccentric anomaly parameter $\eta$, so that a 
uniform distribution in time corresponds to a distribution $f(\eta) = (1 - 
e\cos\eta)/2\pi$.

The orbital periods of G-dwarf stars in the solar neighborhood were found by 
\cite{duquennoy1991} to follow a log-normal distribution with a mean period of 
180 years. \cite{fischer1992} also found a log-normal period distribution for 
M-dwarfs in the solar neighborhood with a mean period similar to that of the 
G-dwarfs.  In terms of logarithm of the period $P$, the distribution found by 
\cite{duquennoy1991} has mean $\mu_{\log P} = 2.23$ and dispersion 
$\sigma_{\log P} = 2.3$ (where $P$ is in years, and the logarithm is base 10).  
We shall use this as the fiducial binary model in this paper, but will also 
allow $\mu_{\log P}$ and $\sigma_{\log P}$ to take on other values.

As with the period distribution, we use distributions of the mass ratio $q$ and 
eccentricity $e$ observed in G-dwarf stars in the solar neighborhood 
(\citealt{duquennoy1991}).  When comparing different empirically derived mass 
functions, \cite{duquennoy1991} showed the distribution of mass ratios $q$ was 
best fit by the Gaussian mass function considered in \cite{kroupa1990} with 
mean $\bar q = 0.23$ and dispersion $\sigma_q = 0.32$.  This is somewhat 
misleading, since for $q > 0.5$ the distribution is in fact consistent with the 
power-law initial mass functions of Salpeter and Kroupa 
(\citealt{salpeter1955}, \citealt{kroupa2001}).  At smaller mass ratios the 
distribution decreases sharply compared to the well-established Kroupa initial 
mass function of single stars, which implies that small mass ratios are 
strongly affected by interaction between primary and secondary during the 
formation of the binary system.

To further complicate matters, 
\cite{mazeh1992} found that short-period binaries tend to have higher mass 
ratios than those found by \cite{duquennoy1991}.  By analyzing spectroscopic 
binaries, they found that binaries with periods shorter than 3000 days have 
mass ratios consistent with a uniform distribution, although the Poisson errors 
in their analysis were quite large.  Subsequent studies 
(\citealt{goldberg2003}, \citealt{halbwachs2004}) have shown that the mass 
ratio distribution in short-period binaries is bimodal (also seen in previous 
samples; cf. \citealt{trimble1990}), with a peak at low mass ratios similar to 
that of the long-period binaries but with another peak near $q \approx 1$.  In 
the sample analyzed by \cite{goldberg2003}, the peak at high mass ratios is 
smaller for primaries with masses larger than 0.6 $M_\odot$, but larger for 
halo stars.  In view of lingering uncertainties in the nature of the mass ratio 
distribution, for simplicity we will adopt the Gaussian distribution from 
\cite{duquennoy1991} for long-period binaries ($P > 3000$ days) and a uniform 
distribution for short-period binaries ($P < 3000$ days).  For the distribution 
of the primary mass $f(m)$ we will use the Kroupa initial mass function 
corrected for binaries (\citealt{kroupa2002}). 

The distribution of eccentricities $f(e|\log P)$ was found by 
\cite{duquennoy1991} to have three different regimes depending on the period.  
For periods of 11 days or shorter, the orbits are circularized due to tidal 
forces and are therefore approximated to have $e = 0$. For periods between 11 
days and 1000 days, $f(e)$ can be approximated by a Gaussian with mean $\bar e 
= 0.25$ and dispersion $\sigma_e = 0.12$.  For periods longer than 1000 days, 
higher eccentricities are more prevalent and the distribution approximately 
follows $f(e) = \frac{3}{2}e^{1/2}$. Among these, the Gaussian regime (11 days 
$< P <$ 1000 days) has the greatest impact on the velocity distribution at 
velocities of order km/s.  While the adopted distributions of mass ratios and 
eccentricities are undoubtedly only a rough approximation to the true 
distributions, our central results (presented in section \ref{sec:correcting_dispersion}) 
will prove to be quite insensitive to the nature of the adopted distributions.

An important effect that must be taken into account is the effect of mass 
transfer between the stars if the primary star is a red giant whose size is 
larger than the radius of its Roche lobe (\citealt{paczynski1971}). In such a 
case, matter from the surface of the giant will accrete onto the smaller star 
and the separation between the stars will decrease. The end result is that 
either the other star will explode in a supernova Ia, or the stars will 
eventually merge.  This effect is not included in the distributions of 
\cite{duquennoy1991} because their sample consisted entirely of dwarf stars.  
Therefore, we make the approximation of excluding systems whose primary star is 
larger than its Roche lobe, assuming the binary to be destroyed over a 
timescale much less than 1 Gyr.  We use an approximation to the radius 
$r_L(a,q)$ of the Roche lobe given by \cite{eggleton1983}. While this radius is 
not exactly correct for eccentric orbits, a recent smoothed-particle 
hydrodynamics simulation of an eccentric binary of mass ratio $q=0.6$ by 
\cite{church2009} found the Roche lobe radius to decrease only slightly with 
eccentricity. They also derive a fitting function for the Roche lobe radius 
with an eccentricity $e$, given by $r_L(e) = r_L(e=0)(1-0.16e)$.  We find using 
this formula that the velocity distribution changes only by a small amount 
(less than 2\% at 5 km/s) compared to when using the Roche lobe radius 
evaluated at pericenter, given by $r_L(e=0)$ above.  For the following 
calculations, we therefore adopt the Roche lobe evaluated at pericenter for an 
eccentric orbit.

The radius of each star is found by estimating its effective temperature from 
an isochrone of given age $t_{g}$ in the stellar population synthesis model of 
\cite{girardi2004}.  This, together with its magnitude, provides an estimate of 
the stellar radius. We denote $M_V(m;t_{g})$ and $R(m;t_{g})$ as the absolute 
V-band magnitude and radius (respectively) of a star of mass $m$ assigned by an 
isochrone of age $t_g$.  If the star lies on the horizontal branch or 
asymptotic giant branch, instead of using its present radius (which may be 
small) we compare its Roche lobe to the largest radius previously attained by 
the star at the end of its red giant phase.

We shall express our formula in terms of an absolute V-band upper magnitude 
limit $M_{lim}$ and age $t_g$. Assuming the lower magnitude limit to 
be near the tip of the red giant branch, we find the velocity distribution is 
quite insensitive to the exact value of the lower magnitude limit because the 
suppression of binaries due to Roche lobe overflow dominates the 
high-luminosity end of the red giant branch.

\begin{figure}
	\includegraphics[height=1.0\hsize,angle=-90]{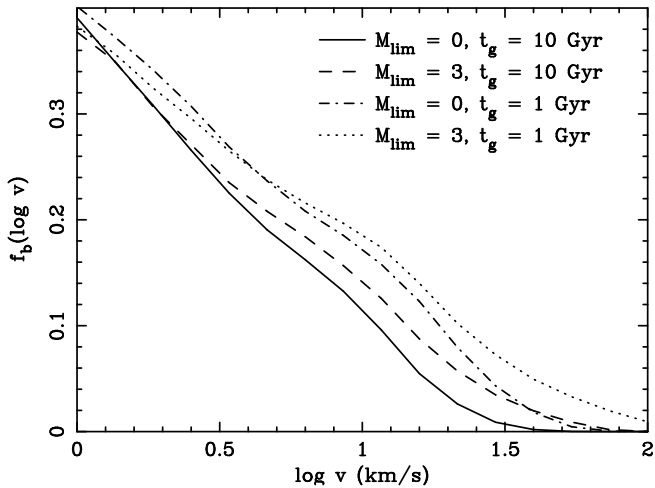}
\caption{Distribution of velocities in the center-of-mass frame of binary 
systems, plotted for different absolute magnitude limits $M_{lim}$ = 0, 3 and 
stellar ages $t_g$ = 1 Gyr, 10 Gyr. The suppression of binaries due to 
Roche-lobe overflow becomes important for velocities $\gtrsim$ 10 km/s. Except 
at the turnover point $v \approx 10$ km/s, the distribution behaves locally as 
a log-normal to good approximation. }
\label{fig:logvhists}
\end{figure}

Averaging over distributions for all the model parameters and dropping the $z$ 
subscript for readability, we obtain

\begin{eqnarray}
\label{eq:fb_logv}
\lefteqn{
f_b(\log |v|; M_{lim}, t_{g}) ~ ~ = } && \\ \nonumber
&& \frac{3}{8\pi^2} \int_{-1}^1 d(\cos \theta) \int_0^{2\pi} d\psi \int_0^1 f(e|\log P)de \int_0^{2\pi}f(\eta)d\eta \\ \nonumber
& \times & \int_0^1 f(q|\log P)dq \int_{0}^{\infty} f(m)dm \cdot 
\Theta\left[M_{lim}-M_V(m;t_{g})\right] \\ \nonumber
%& \times & \Theta\left[r_L(a,q) - R(m;t_{g})\right]\cdot f_P(k - 3 \log|v|),
& \times & \Theta\left[r_L(a,q) - R(m;t_{g})\right] \frac{\exp\left\{\frac{-\left[3 \log|v| - k + \mu_{\log P}\right]^2}{2\sigma_{\log P}^2}\right\}}{\sqrt{2\pi\sigma_{\log P}^2}},
\end{eqnarray}

where $\Theta[x]$ is the Heaviside step function. The variable $k$ is a 
function of all the other parameters according to (\ref{k_formula}), with $\phi 
\equiv \phi(\eta)$. We use a Monte Carlo simulation to perform the integration 
over a grid of $\log |v|$ values and interpolate to find $f_b(\log |v|)$. In 
fig.~\ref{fig:logvhists} we plot $f_b(\log |v|)$ for different ages and 
absolute magnitude limits. This figure shows that for velocities $\gtrsim 10$ 
km/s, suppression of binaries due to Roche-lobe overflow becomes important.

Asymptotically for large velocities, the velocity distribution in 
eq.~\ref{eq:fb_logv} behaves as a log-normal with dispersion similar to that of 
the period distribution, $\sigma_{\log P}$. This can be seen as follows: first, 
at large velocities, the log-normal in the integrand is far from its maximum 
and therefore varies slowly in the model parameters (eq.~\ref{k_formula}) 
compared to their respective probability distributions, provided $k$ is not 
large and negative.  Therefore as a first approximation we can apply the method 
of steepest descents and find the resulting distribution to be log-normal 
with a mean given by $\log |\bar v| = \frac{1}{3}\left(\bar k - \mu_{\log 
P}\right)$. However, the approximation of slowly varying log-normal is not 
strictly true, since $k$ becomes large and negative if the mass ratio $q$ and 
direction function $g_e(\theta,\psi,\phi)$ are close to zero. This means that 
the mean $\bar k$ is in fact a function of $v$. We find, however, that locally 
$\bar k$ is linear in $\log |v|$ to good approximation, with the result that 
the velocity distribution still behaves locally as a log-normal but with a 
somewhat different dispersion from $\sigma_{\log P}$. We will use this to 
construct a fitting function for the binary velocity distribution in section 
\ref{sec:fitting_function}.

\section{Velocity distribution of a population of binary 
stars}\label{sec:likelihoods}

\subsection{Single- and two-epoch likelihood functions}\label{subsec:likelihoods}

Suppose that among a population of stars, a fraction $B$ of them are in binary 
systems.  Further suppose that the velocity distribution for stars not in 
binary systems is Gaussian with dispersion $\sigma_0$ and systemic velocity 
$\bar v$.  The velocity likelihood function will have the following form:

\begin{equation}
\label{f_v}
L(v|B,\sigma_0,\bar v) ~ = ~ (1 - B) \frac{e^{-(v-\bar 
v)^2/2\sigma_0^2}}{\sqrt{2\pi\sigma_0^2}} ~ + ~ B L_b(v|\sigma_0,\bar v)
\end{equation}
where $L_b(v|\sigma_0,\bar v)$ is the likelihood for binary stars. To derive 
the binary likelihood, we note that the component of the velocity \emph{not} 
due to the binary orbit is $v_{nb} = v - v'$, where $v'$ is the line-of-sight 
component of the velocity in the center-of-mass frame of the binary system. To 
find the binary likelihood we therefore average the velocity distribution in 
$v_{nb}$ over the distribution of the binary component $v'$ given in 
eq.~\ref{eq:fb_logv}: 

\begin{equation}
\label{fv_binary_part}
L_{b}(v|\sigma_0,\bar v) ~ = ~ \int_{-\infty}^{\infty}\frac{e^{-(v-v'-\bar v)^2/2\sigma_0^2}}{\sqrt{2\pi\sigma_0^2}}\frac{f_b(\log|v'|)}{2|v'|\ln 10}dv'
\end{equation}

%L_{b}(v|\sigma_0,\bar v) ~ ~ = ~ ~ } \\
%&& \int_{-\infty}^{\infty}\frac{1}{2}\left\{\frac{e^{-(v-|v'|-\bar v)^2/2\sigma_0^2} + e^{-(v+|v'|-\bar v)^2/2\sigma_0^2}}{\sqrt{2\pi\sigma_0^2}}\right\} f_b(\log|v'|)d(\log |v'|) \nonumber

%The two terms in the integrand correspond to $v' = +|v'|$ and $v' = -|v'|$, 
%respectively.

The factor of 2 in the denominator arises from the fact that $f_b(\log|v|)$ is normalized in $\log |v|$, whereas the likelihood is normalized in $v$ (allowing for positive and negative velocities).

By taking the second moment of the velocity distribution in eq.~\ref{f_v}, one 
obtains the result that

\begin{equation}
\label{dispersion_total}
\sigma^2 = \sigma_0^2 + B \sigma_b^2
\end{equation}
where $\sigma$ is the measured dispersion and $\sigma_b$ is the binary 
dispersion found by taking the second moment of the binary velocity 
distribution $f_b(v) = f_b(\log |v|)/2|v|\ln 10$.  As in the usual case, it can 
be shown that given a normally distributed measurement error with dispersion 
$\sigma_m$, one need only make the replacement $\sigma_0^2 \longrightarrow 
\sigma_0^2 + \sigma_m^2$ in the above formulas.

Next, it is desirable to have a likelihood function for velocities measured at 
two different epochs. Since velocity changes of order km/s over a timescale of 
years is entirely negligible for nonbinary stars, the most fruitful approach is 
to use a likelihood in the difference $\Delta v$ between the two velocities.  
Keeping in mind the log-normal behavior of the velocity distribution, we write 
the binary part of the likelihood as $g_b(\log|\Delta v|;\Delta t)$. As with 
the single-epoch velocity distribution $f_b(\log|v|)$, we use a Monte Carlo 
simulation to calculate $g_b(\log|\Delta v|;\Delta t)$.  For each binary in the 
simulation we find $\Delta v$ by evolving the orbital phase to its value after 
a time $\Delta t$ and calculating the resulting change in velocity.  In the 
absence of measurement error, the nonbinaries will have zero change in 
velocity, so the total likelihood can be written as

\begin{equation}
\label{eq:f_dv_noerr}
L(\Delta v|\Delta t,B) = (1-B)\delta(\Delta v) + B\frac{g_b(\log|\Delta 
v|;\Delta t)}{2|\Delta v| \ln 10}
\end{equation}

If there is a normally distributed measurement error, the likelihood must be 
averaged over two Gaussians of widths $\sigma_{m,1}$ and $\sigma_{m,2}$ for the 
first and second velocity errors, respectively. A little calculation shows this 
to be equivalent to averaging over a single Gaussian with dispersion 
$\sigma_{2e}$, which is the equivalent 2-epoch measurement error:

\begin{equation}
\label{eq:sigma_2m}
\sigma_{2e}^2 = \sigma_{m,1}^2 + \sigma_{m,2}^2,
\end{equation}

\begin{eqnarray}
\label{eq:f_dv}
\lefteqn{
L(\Delta v|\Delta t,B,\sigma_{2e}) = (1 - B) \frac{e^{-\Delta v^2/2\sigma_{2e}^2}}{\sqrt{2\pi\sigma_{2e}^2}} } \\
& + & B\int_{-\infty}^{\infty}\frac{e^{-(\Delta v-\Delta 
v')^2/2\sigma_{2e}^2}}{\sqrt{2\pi\sigma_{2e}^2}} \frac{g_b(\log|\Delta 
v'|;\Delta t)}{2|\Delta v'| \ln 10}d(\Delta v') \nonumber
\end{eqnarray}

Note that the likelihood is identical in form to that of eqs.~\ref{f_v} and 
\ref{fv_binary_part}, since in both cases we are averaging the distribution 
over a Gaussian. Both the single- and multi-epoch likelihoods will be put to 
use in later sections.

\subsection{Threshold fraction of a binary population}\label{sec:threshold_fraction}

A convenient observable quantity for characterizing a binary population is the 
\emph{threshold fraction}, defined as the fraction $F$ of stars in a sample 
which exhibit a change in radial velocity greater than a threshold $\Delta v$ 
after a time $\Delta t$ between measurements.  For $\Delta v > 1$ km/s, this 
fraction is typically smaller than 0.2, so the threshold \emph{number} (given 
by $n = NF$ where $N$ is the number of stars) follows a Poisson distribution 
with mean $\bar n = N\bar F$.  Therefore the distribution of $F$ is 
characterized by a single number, the mean threshold fraction $\bar F$, and the 
expected error can be estimated.  In particular, the error in $F$ is 
approximately $\sqrt{\bar F/N}$ (appendix \ref{sec:appendix_b}). For notational 
simplicity, for the remainder of this paper we will refer to the mean threshold 
fraction $\bar F$ as simply the threshold fraction $F$ (without the bar), with 
the understanding that the observed threshold fraction will have a Poisson 
scatter about this value.

Despite its straightforward definition, there are two difficulties in measuring 
the threshold fraction from actual data sets. First, often there does not exist 
a common time interval $\Delta t$ between measurements in the sample, but 
rather several time intervals for various subsets of stars. Furthermore, 
different velocity measurements have their own associated measurement errors 
and this in turn affects the measured value of $F$. The latter issue can be 
dealt with in an approximate way by using the median (or other robust location 
estimator) of the measurement error of the sample, in terms of which 
$\bar\sigma_{2e} = \bar\sigma_m\sqrt{2}$ (eq.~\ref{eq:sigma_2m}).  However, 
both problems can be surmounted more rigorously by estimating the error-free 
threshold fraction $F_0$ via a Bayesian or maximum-likelihood approach.  By 
using the likelihood in $\Delta v$ defined in eq.~\ref{eq:f_dv}, the threshold 
fraction at a particular threshold and time interval can be estimated even if 
measurements were taken at various epochs---moreover, the inferred threshold 
fraction $F_0$ is free of measurement error.  This method will be demonstrated 
in section \ref{sec:multi_epoch}.

The threshold fraction without measurement error, which we denote by $F_0$, can 
be expressed in terms of the binary two-epoch velocity distribution $g_b(\log 
|\Delta v|;\Delta t)$ by taking the integral of eq.~\ref{eq:f_dv_noerr} with 
respect to $|\Delta v'|$ from a threshold $\Delta v$ to $\infty$:

\begin{equation}
\label{eq:F_dv_noerr}
F_0(\Delta v|\Delta t,B) = B\int_{\Delta v}^{\infty} \frac{g_b(\log|\Delta 
v'|;\Delta t)}{|\Delta v'|\ln 10}d|\Delta v'| \nonumber
\end{equation}

Note that in the absence of measurement error, the threshold fraction $F_0$ 
scales linearly with the binary fraction $B$. The threshold fraction with 
measurement error is likewise obtained by taking the integral of 
eq.~\ref{eq:f_dv} from $\Delta v$ to $\infty$, with the result

\begin{eqnarray}
\label{eq:F_dv}
\lefteqn{
F(\Delta v|\Delta t,B,\sigma_{2e}) = (1 - B)\textrm{erfc}\left[\frac{\Delta 
v}{\sqrt{2}\sigma_{2e}}\right] } \\
& + & B\int_{-\infty}^{\infty}\textrm{erfc}\left[\frac{\Delta v-\Delta 
v'}{\sqrt{2}\sigma_{2e}}\right] \frac{g_b(\log|\Delta v'|;\Delta t)}{2|\Delta 
v'|\ln 10}d(\Delta v') \nonumber
\end{eqnarray}
where $\sigma_{2e}$ is the 2-epoch measurement error given by 
eq.~\ref{eq:sigma_2m}. Note that in the limit as $\sigma_{2e} \rightarrow 0$, 
the first term goes to zero while the complementary error function in the 
integrand reduces to a step function 2$\Theta(\Delta v' - \Delta v)$, so that 
eq.~\ref{eq:F_dv} reduces to eq.~\ref{eq:F_dv_noerr} as expected.

\section{Constraining properties of a binary population by multi-epoch 
observations}\label{sec:multi_epoch}

In this section we investigate how properties of a population of binary stars 
affect the observed velocity distribution measured at two or more epochs.  
Specifically, we explore how our proposed observable, the threshold faction $F$ 
(section \ref{sec:threshold_fraction}), will be affected by changes in the 
underlying binary fraction $B$, absolute magnitude limit, stellar age, size of 
the measurement error, and time interval between measurements.  We will also 
demonstrate how the binary fraction $B$ can be inferred by a likelihood 
analysis, and show how this leads to a better determination of the threshold 
fraction $F$.  We first consider binary models with our fiducial period 
distribution (inferred from the solar neighborhood) and then explore how the 
inferred binary fraction is affected if the assumed period distribution 
parameters are incorrect.  Unfortunately, and as we discuss more fully in the 
next section, the effect of changing the binary fraction $B$ on the observed 
binary velocities can be mimicked closely by altering the assumed distribution 
of orbital periods (i.e. changing the parameters $\mu_{\log P}$, $\sigma_{\log 
P}$). While this is bad news for any attempt at constraining the underlying 
properties of a galaxy's binary population in full generality, it turns out to 
be good news for correcting the observed velocity dispersion for the effects of 
binary orbital motion, as we will show in section 
\ref{sec:correcting_dispersion}.

%The simplest likelihood analysis would take the binary fraction $B$ as a free 
%model parameter in the likelihood. In a more general analysis, one would take 
%as free parameters not only the binary fraction, but also the period 
%distribution parameters $\mu_{\log P}$ and $\sigma_{\log P}$. Then all three 
%parameters would be marginalized over to find the most general constraints.  
%Unfortunately, neither the data nor simulations can give us a prior in these 
%parameters.  Further, as we will show in section 
%\ref{sec:correcting_dispersion}, the period distribution parameters are nearly 
%degenerate with binary fraction over the scale of km/s, so the constraints on 
%each individual parameter would be quite poor. In section 
%%\ref{sec:period_dist} 
%we will investigate how many stars would be required to break this degeneracy; 
%for now, we will fix the period distribution parameters and take only the 
%binary fraction $B$ as a free model parameter.

First, let us make the rather optimistic assumption that the distribution of 
binary orbital periods is approximately universal, so that it follows our 
fiducial choice $\mu_{\log P}=2.23$, $\sigma_{\log P}=2.3$ (section 
\ref{sec:derivations}).  Before launching into the full-fledged calculation, 
one would like to estimate how well the fiducial binary fraction $B$ can be 
constrained for a given sample, or conversely, how many stars are required to 
constrain $B$ by a certain amount.  To simplify matters, let us assume we have 
a data set where the two epochs have the same time interval $\Delta t$ between 
them and the same measurement error $\sigma_m$. The equivalent two-epoch 
measurement error is then $\sigma_{2e} = \sigma_m \sqrt{2}$ 
(eq.~\ref{eq:sigma_2m}).

\begin{figure}
\includegraphics[height=1.0\hsize,angle=-90]{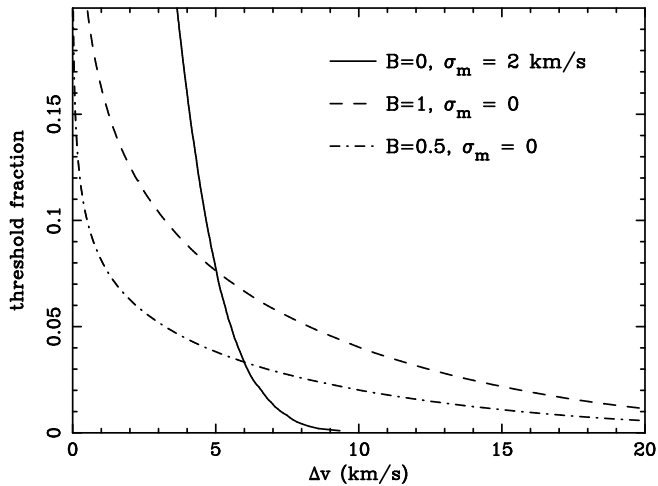}
\caption{Threshold fraction $F(\Delta v|\Delta t,B,\sigma_m)$, defined as the 
fraction of stars with observed change in velocity greater than a threshold 
$\Delta v$ after a time interval $\Delta t$ between measurements.  The solid 
curve has no binaries ($B=0$) and a measurement error $\sigma_m = 2$ km/s. The 
other curves are plotted from a Monte Carlo simulation for binary fractions $B$ 
= 1 and 0.5, with $\Delta t$ = 1 year and no measurement error.  The stellar 
population has an age $t_g$ = 10 Gyr and the absolute magnitude limit $M_{lim}$ 
= 1.  Given a measurement error $\sigma_m$, the binary fraction can be 
constrained for thresholds $\Delta v \gtrsim \Delta v_{tail}$, where $\Delta 
v_{tail}$ is the point of intersection where $F(\Delta v|\Delta t,B=0,\sigma_m) 
= F(\Delta v|\Delta t,B,\sigma_m=0)$. For a given binary fraction $B$, the 
total threshold fraction without measurement error is given by $B \times 
F(\Delta v|\Delta t, B=1, \sigma_m=0)$.  }
\label{fig:dvchist}
\end{figure}

First, consider the mean threshold fraction of the binaries without measurement 
error, denoted by $F_0(\Delta v|\Delta t,B=1)$ (eq.~\ref{eq:F_dv_noerr}).  In 
that case the threshold fraction scales with the binary fraction $B$, i.e.  is 
given by $F_0(\Delta v|\Delta t,B) = B F_0(\Delta v|\Delta t,B=1)$.  Now 
consider the threshold fraction without binaries, but with a measurement error 
$\sigma_m$.  The two-epoch measurement error is then $\sigma_{2e} \approx 
\sqrt{2}\sigma_m$ (eq.~\ref{eq:sigma_2m}), and the threshold fraction $F(\Delta 
v|\Delta t,B=0,\sigma_{2e})$ is given by the complementary error function 
(first term in eq.~\ref{eq:F_dv} with $B=0$).  In fig.~\ref{fig:dvchist} we 
plot the threshold fraction $F_0(\Delta v|\Delta t,B)$ produced by a Monte 
Carlo simulation with an absolute magnitude limit $M_{lim}=1$ and stellar age 
$t_g$ = 10 Gyr.  The threshold fraction is plotted for different binary 
fractions, and we also plot the threshold fraction from measurement error with 
$\sigma_m = 2$ km/s.  Near the point of intersection $\Delta v_{tail}$ where 
$F_0(\Delta v_{tail}|\Delta t,B) = F(\Delta v_{tail}|\Delta 
t,B=0,\sigma_{2e})$, the effect of binary stars becomes noticeable over the 
measurement error.  Since the Poisson errors are larger at higher velocity 
thresholds, to first approximation we can say that the error-free threshold 
fraction $F_0$ is best constrained at thresholds near $\Delta v_{tail}$.  It 
follows that the fiducial binary fraction will be constrained by the stars with 
$\Delta v \gtrsim \Delta v_{tail}$.  (For a rough approximation, one can also 
use $\Delta v_{tail} \approx 2\sigma_{2e} \approx 2\sqrt{2}\sigma_m$.) A little 
algebra (see appendix \ref{sec:appendix_b}) shows that to constrain the binary 
fraction to within a fractional accuracy of $\epsilon_b$, the number of stars 
required is approximately

\begin{equation}
N(\epsilon_b) \approx \frac{1}{\bar F(\Delta v_{tail})} 
\left(2B\over\epsilon_b\right)^2
\label{eq:N_sigB}
\end{equation}

\begin{figure}
	\includegraphics[height=1.0\hsize,angle=-90]{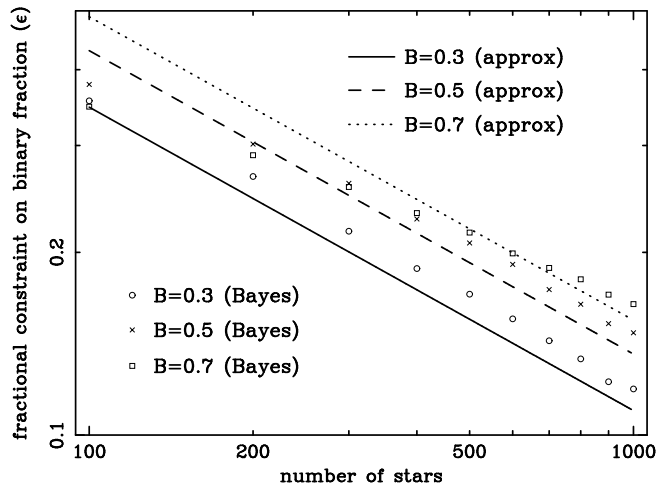}
\caption{Fractional constraint $\epsilon_b$ on the binary fraction $B$, defined 
by eq.~\ref{eq:N_sigB}. In this plot we a measurement error $\sigma_m = 2.0$ 
km/s.  For comparison we also plot the 95\% confidence interval obtained by a 
Bayesian analysis of simulated two-epoch data from a random sample of $N$ 
stars, averaged over a hundred realizations. A uniform prior is assumed for 
$B$.}
\label{fig:approx}
\end{figure}

In fig.~\ref{fig:approx} we graph the approximation formula for different 
values of $B$ and compare to the 95\% confidence interval in the binary 
fraction inferred by a Bayesian analysis of the simulated data (described later 
in this section).  As is evident for the $B=0.7$ curve, the approximation 
formula differs for high binary fractions because $B > 1$ is not allowed in the 
Bayesian analysis.  The approximation formula is discussed further in appendix 
\ref{sec:appendix_b}.

\begin{figure}
	\includegraphics[height=1.0\hsize,angle=-90]{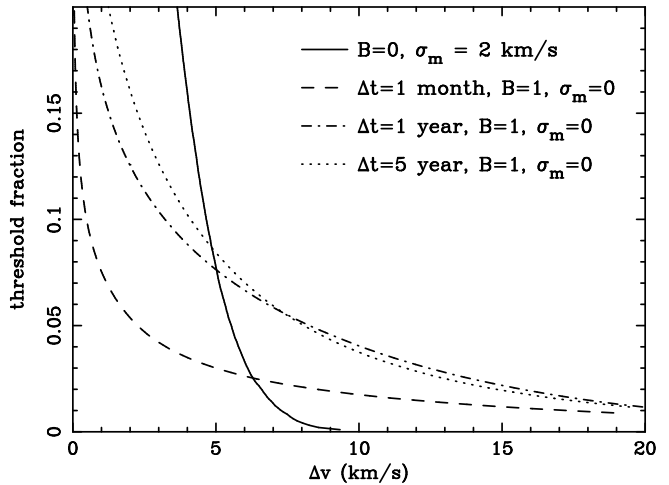}
\caption{Threshold fraction $F(\Delta v|\Delta t,B,\sigma_m)$, defined as the 
fraction of stars with observed change in velocity greater than a threshold 
$\Delta v$ after a time interval $\Delta t$ between measurements.  The solid 
curve has no binaries ($B=0$) and a measurement error $\sigma_m = 2$ km/s. The 
other curves are plotted from a Monte Carlo simulation for different time 
intervals $\Delta t$, with no measurement error and a binary fraction $B=1$.  
The stellar population has an age $t_g$ = 10 Gyr and the absolute magnitude 
limit $M_{lim}$ = 1.  Given a measurement error $\sigma_m$, the binary 
fraction can be constrained for thresholds $\Delta v \gtrsim \Delta v_{tail}$, 
where $\Delta v_{tail}$ is the point of intersection where $F(\Delta v|\Delta 
t,B=0,\sigma_m) = F(\Delta v|\Delta t,B,\sigma_m=0)$. For a given binary 
fraction $B$, the total threshold fraction without measurement error is given 
by $B \times F(\Delta v|\Delta t, B=1, \sigma_m=0)$.  }
\label{fig:dvchist_dt}
\end{figure}

In fig.~\ref{fig:dvchist_dt} we plot the threshold fraction $\bar F(\Delta 
v|\Delta t,B=1,\sigma_{2e}=0)$ produced by a Monte Carlo simulation with an 
absolute magnitude limit $M_{lim}=1$ and stellar age $t_g = 10$ Gyr for 
different time intervals $\Delta t$.  We find that for a measurement error 
$\sigma_m = 2$ km/s, the observable threshold fraction steadily increases as 
$\Delta t$ is increased, until roughly $\Delta t = 1$ year. This result depends 
somewhat on the mass ratio distribution, since higher mass ratios result in 
higher velocities for a given orbital period. If the mass ratio distribution in 
\cite{duquennoy1991} is assumed for all periods (as opposed to the uniform 
distribution we adopt for $P < 3000$ days), the observable threshold fraction 
increases until roughly $\Delta t = 2$ years. In any case, unless the 
measurement error is smaller than 2 km/s, little is gained by extending the 
interval from 1-2 years to 5 or more years. 

\begin{figure}
	\includegraphics[height=1.0\hsize,angle=-90]{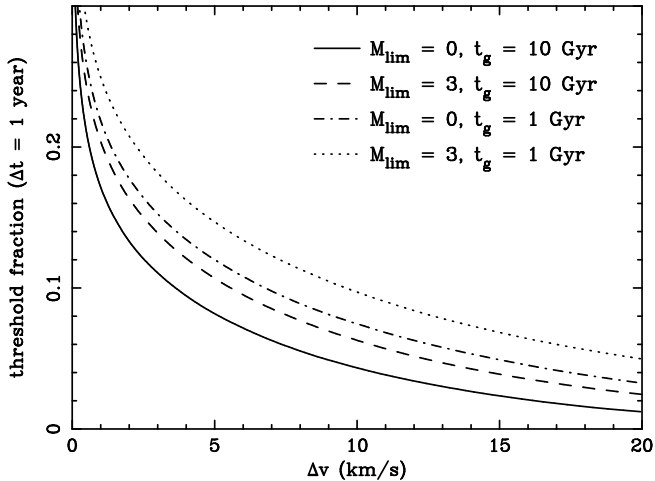}
\caption{Threshold fraction $F(\Delta v|\Delta t,B,\sigma_m)$, defined as the 
fraction of stars with observed change in velocity greater than a threshold 
$\Delta v$ after a time interval $\Delta t$ year between measurements. Here we 
have picked $\Delta t$ = 1 year, binary fraction $B=1$, and no measurement 
error ($\sigma_m=0$). We plot the threshold fraction for different absolute 
magnitude limits $M_{lim}$ = 0, 3 and stellar ages $t_g$ = 1 Gyr, 10 
Gyr.  For a given binary fraction $B$, the total threshold fraction is given by 
$F(\Delta v|\Delta t,B,\sigma_m=0) = B \times F(\Delta v|\Delta 
t,B=1,\sigma_m=0)$.}
\label{fig:dvchist_age}
\end{figure}

In fig.~\ref{fig:dvchist_age} we plot the threshold fraction for different 
absolute magnitude limits $M_{lim}$ = 0, 3 and stellar ages $t_g$ = 1 
Gyr and 10 Gyr. Extending the magnitude limit to fainter magnitudes increases 
the threshold fraction because there is a greater contribution from smaller 
stars with less binary suppression due to Roche-lobe overflow. The threshold 
fraction is also higher for a younger stellar population because of their 
larger mass at a given stage of stellar evolution, which produces higher 
orbital velocities. We find that the threshold fraction changes little for ages 
between 2-10 Gyr; as the stellar age $t_g$ is reduced from 2 Gyr, however, the 
threshold fraction steadily increases.

\begin{figure}
	\includegraphics[height=1.0\hsize,angle=-90]{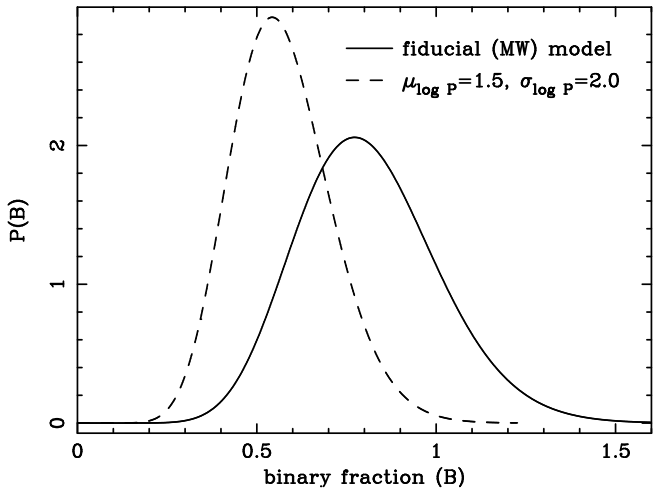}
\caption{Posterior probability distribution of the binary fraction $B$ of a 
simulated galaxy with binary fraction $B=0.5$ and with a period distribution 
characterized by $\mu_{\log P}=1.5$, $\sigma_{\log P}=2$ ($P$ in years).  The 
simulated data sample consisted of 300 stars, each with two velocity 
measurements taken $\Delta t_{data}$ = 2 years apart. Solid curve is the 
posterior calculated assuming the fiducial (solar neighborhood) model, which is 
incorrect for this galaxy. Dashed curve uses the correct model, with the same 
period distribution parameters $\mu_{\log P}$, $\sigma_{\log P}$ given above.}
\label{fig:bf_2models}
\end{figure}

\begin{figure}
	\includegraphics[height=1.0\hsize,angle=-90]{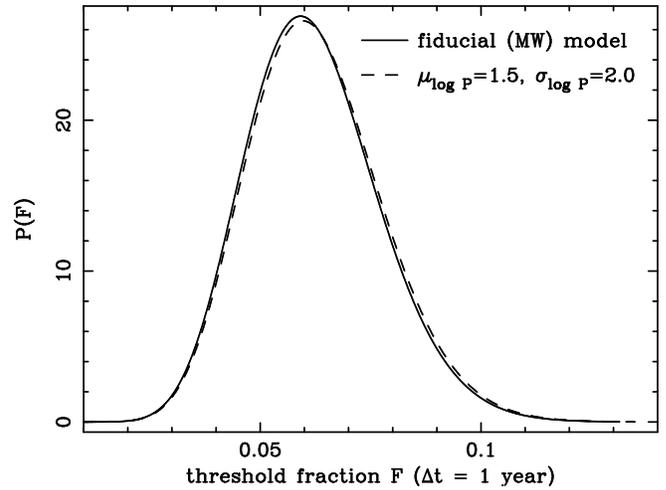}
\caption{Posterior probability distribution of the threshold fraction $F$ of a 
simulated galaxy with binary fraction $B=0.5$ and with a period distribution 
characterized by $\mu_{\log P}=1.5$, $\sigma_{\log P}=2$ ($P$ in years).  The 
threshold fraction $F(\Delta v|\Delta t)$ is defined as the fraction of stars 
with observed change in velocity greater than a threshold $\Delta v$ after a 
time interval $\Delta t$ year between measurements. Here we chose a threshold 
$\Delta v$ = 5 km/s and time interval $\Delta t$ = 1 year; the correct 
threshold fraction for this galaxy is $F \approx 0.05$. The simulated data 
sample consisted of 300 stars, each with two velocity measurements taken 
$\Delta t_{data}$ = 2 years apart, and a measurement error of 2 km/s.  Solid 
curve is the posterior calculated assuming the fiducial (solar neighborhood) 
model, which is incorrect for this galaxy. Dashed curve uses the correct model, 
with the same period distribution parameters $\mu_{\log P}$, $\sigma_{\log P}$ 
given above. Note that the correct threshold fraction can be recovered even if 
the wrong model is assumed (in this case, the fiducial model). }
\label{fig:f_2models}
\end{figure}

To estimate the binary fraction in a two-epoch sample, we use the likelihood 
function $L(\Delta v|\Delta t,B,\sigma_{2e})$ (eq.~\ref{eq:f_dv}).  For the 
sake of illustration, we analyze a simulated galaxy with binary fraction 
$B=0.5$ but with a different period distribution from that of the solar 
neighborhood. We choose the period distribution parameters $\mu_{\log P}=1.5$, 
$\sigma_{\log P}=2$ ($P$ in years) for this galaxy. The simulated data sample 
consists of 300 stars, each with two velocity measurements taken $\Delta 
t_{data}$ = 2 years apart, and a measurement error of 2 km/s.  First we assume 
the fiducial model (with $\mu_{\log P}=2.23$, $\sigma_{\log P}=2.3$, which is 
incorrect for this galaxy) and, assuming a uniform prior in the binary fraction 
$B$, generate a posterior in the binary fraction.  We then repeat this 
procedure using the correct period distribution parameters $\mu_{\log P}$, 
$\sigma_{\log P}$ in our model, whose values are given above.  The resulting 
posteriors are plotted in fig.~\ref{fig:bf_2models}. This figure shows that the 
binary fraction $B$ is a highly model-dependent quantity, and given the unknown 
nature of the period distribution of binaries outside the solar neighborhood, 
the inferred binary fraction must be taken with a grain of salt. However, the 
fiducial binary fraction can still be used as a relative indicator of the 
fraction of observable binaries, as long as it is interpreted in reference to 
the fraction observed in a binary population following the fiducial (solar 
neighborhood) distributions of orbital parameters.

Although the inferred binary fraction $B$ is very model-dependent, this 
analysis is still useful in that it leads to a better determination of the 
threshold fraction $F$, which is more directly observable than the binary 
fraction.  To see this, we use the Monte Carlo to generate the binary threshold 
fraction $F_b$ of each model, for a threshold $\Delta v$ = 5 km/s, time 
interval $\Delta t$ = 1 year, and zero measurement error, i.e.  $\sigma_{2e}$=0 
(see eq.~\ref{eq:F_dv}). We then transform each posterior in 
fig.~\ref{fig:bf_2models} from $B$ to the threshold fraction $F = B F_b$.  The 
renormalized posteriors $P(F)$ are plotted in fig.~\ref{fig:f_2models}; the 
correct threshold fraction for this galaxy is $F \approx 0.05$.  Note that the 
correct threshold fraction can be recovered even if the wrong model is assumed 
(in this case, the fiducial model). 

Since even the stars with velocities smaller than the 
threshold $\Delta v$ are used in the likelihood analysis, the error in the 
threshold fraction $F$ is smaller than if $F$ is measured directly, especially 
for higher velocity thresholds. The approximate error in the threshold fraction 
$F$ estimated by this technique is derived in appendix \ref{sec:appendix_b} and 
given by eq.~\ref{sigF_best_fit}.  Furthermore, the threshold fraction at 
$\Delta t$ = 1 year is recovered even though the data was taken with a time 
interval of $\Delta t_{data}$ = 2 years.  More generally, the threshold 
fraction for a specific time interval can be recovered by the likelihood 
analysis even if the data is taken at various different epochs and with 
different measurement errors. As we will show in section 
\ref{sec:correcting_dispersion}, the threshold fraction can be used to correct 
the measured velocity dispersion of a sample for the effect of binary motion.

%Rather than employing an analytic formula for this (whose form is quite 
%complicated and inefficient to calculate), we use the ``brute force'' method 
%of generating the distribution by binning a large number of points from a 
%Monte Carlo simulation.  The distribution of $f_b(\log|\Delta v'|;\Delta t)$, 
%where $\Delta v'$ is the change in binary velocity not including the 
%measurement error, is plotted in fig.~\ref{fig:logdvhist} for different values 
%of $\Delta t$. In these graphs, the peak at high velocities corresponds to 
%binary stars that have completed one or more orbital cycles in the given time 
%interval, while the tail at low velocities are binaries which have only 
%traversed a small fraction of their orbit.

%Over the range $-3 < \log|\Delta v'| < 1$ the distribution varies only by a 
%small amount; this means that $f(|\Delta v'|) \propto 1/|\Delta v'|$ over this 
%range.  This apparent divergence is softened, however, for very small values of 
%$\Delta v'$. The exact behavior of $f(\Delta v')$ for such small values is 
%unimportant because any $\Delta v \lnsim \sigma_m$ will be swamped by the 
%measurement error. What is important is the area beyond $\Delta v' \approx 
%\sigma_m$ (more precisely, $\Delta v_{tail}$ discussed earlier) which gives the 
%fraction of binaries with an observable change in velocity beyond the expected 
%measurement error.  These are the stars that enable us to constrain the binary 
%fraction.

\begin{figure}
	\includegraphics[height=1.0\hsize,angle=-90]{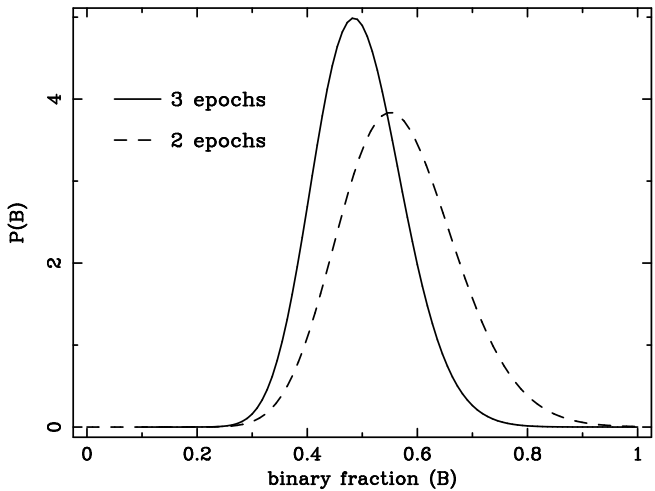}
\caption{Posterior probability distribution for binary fraction of a simulated galaxy with 500 stars and binary fraction $B=0.5$. Solid curve is calculated from three velocity measurements, whereas dotted curve uses only the first two velocity measurements. $v_1$ and $v_2$ were taken 1 year apart, while $v_2$ and $v_3$ were taken 10 years apart.}
\label{fig:bf_3epoch}
\end{figure}

Finally, one may naturally wonder: how much are the constraints improved by 
including more than two epochs in the analysis?  To address this question, we 
do a similar calculation on simulated 3-epoch data using the Monte Carlo to 
generate a 3-epoch likelihood, $L(\Delta v_{21}, \Delta v_{32}|\Delta 
t_{21},\Delta t_{32},B,\sigma_{3e})$, where the indices refer to three velocity 
measurements $v_1$, $v_2$, and $v_3$, and $\sigma_{3e}$ is the 3-epoch 
measurement error defined analogously to eq.~\ref{eq:sigma_2m}. We generate a 
data sample of 500 stars from a simulated galaxy with binary fraction $B=0.5$ 
and the period distribution parameters taking their fiducial values. For 
comparison, we generate a posterior $P(B)$ from the two-epoch calculation which 
ignores $v_3$. The results are plotted in fig.~\ref{fig:bf_3epoch}.  The 
velocity measurements $v_1$ and $v_2$ were taken one year apart, while $v_2$ 
and $v_3$ were taken ten years apart. While the most probable inferred binary 
fraction $B_{fit}$ did improve in this case, the 95\% confidence limits are 
only improved by $\approx 25\%$. The fractional improvement in the confidence 
limits is even less for smaller data sets; this is because in a sample of a few 
hundred stars, there is significant scatter in the binary fraction, and the 
inferred binary fraction $B_{fit}$ has in fact a significant probability of 
becoming worse when more epochs are added to the sample.  To constrain the 
binary and/or threshold fraction, we therefore find it a more profitable 
strategy to make two-epoch measurements over a larger sample of stars, as 
opposed to adding more repeat measurements over an existing sample (assuming a 
similar number of overall measurements in either case).

A possible complicating factor in the above analysis is that selection criteria 
for making repeat measurements can bias the inferred threshold fraction and 
binary fraction. If stars whose spectra yield multiple peaks in the cross 
correlation function are singled out for repeat measurements, the multi-epoch 
sample may have an inordinately high binary fraction compared to the overall 
stellar population.  This selection bias is probably not significant in red 
giant stars due to the typically large difference in luminosity between primary 
and secondary star.  However, in samples that contain a significant fraction of 
main sequence stars the bias may be more problematic, although an upper bound 
on the binary dispersion can still be obtained.

\section{Correcting the observed velocity dispersion from multi-epoch 
data}\label{sec:correcting_dispersion}

%If radial velocities are measured at two or more epochs, there are two 
%fundamentally different approaches toward correcting the dispersion for binary 
%motion.  The first approach is to single out stars with a large change in 
%velocity compared to the measurement error and remove them from the sample, 
%since these have a high probability of inflating the dispersion (cf.  
%\citealt{olszewski1996}). While this can reduce the amount by which the 
%dispersion is inflated by binaries, it does little toward constraining the 
%binary dispersion and indeed makes it difficult to determine the effect of 
%binaries on the dispersion at smaller velocities. However, in small data sets 
%(less than 100 stars) this approach may be necessary as the binary dispersion 
%would be poorly constrained in any case.

In the previous section we demonstrated how uncertainties in the underlying 
period distribution can adversely affect our ability to constrain the 
underlying binary fraction from multi-epoch data. Here we demonstrate that if 
our goal is to correct the observed velocity dispersion for the effects of 
binary stars, the degeneracy between the period distribution parameters and 
binary fraction is quite useful: regardless of the precise nature of the binary 
population, its effect on the observable threshold fraction $F$ can be directly 
related to the associated correction in the observed velocity dispersion in a 
model-independent way.

The important degeneracy arises from the log-normal behavior of the binary 
velocity distribution $f_b(\log |v|)$ (eq.~\ref{eq:fb_logv}). Binary orbital 
motion along the line of sight of order km/s is the most important for the 
intrinsic dispersions we are interested in. For these velocities, the value of 
$\log |v|$ is far from the mean of the log-normal, which is approximately $-1$ 
for a magnitude limit $M_{lim}=1$ and age $t_g$ = 10 Gyr.  The exponent of the 
log-normal is approximately linear over the scale of km/s, so we can write it 
as $-\beta - \alpha \ln |v|$.  Therefore the binary part of the velocity 
distribution can be written as $f(v) \propto Be^{-\beta}|v|^{-1-\alpha}$, where 
$B$ is the binary fraction.  If the mean binary period $\mu_{\log P}$ is 
varied, the log-normal is offset in log-space so to good approximation only 
$\beta$ changes; therefore the velocity distribution $f(v)$ changes by a 
constant factor over the scale of km/s.  If the dispersion of the period 
distribution $\sigma_{\log P}$ is varied, both the offset $\beta$ and the slope 
$\alpha$ change; however, the slope changes by a relatively small amount for 
$\sigma_{\log P}$ ranging from 1-3 (its viable range of values; see section 
\ref{sec:period_dist}), so again the velocity distribution changes by an 
approximately constant factor.  The important point is if that the parameters 
$\sigma_{\log P}$ and $\mu_{\log P}$ are varied, they change the velocity 
distribution by an amount which is nearly the same over the scale of several 
km/s---in other words, \emph{they behave similarly as if the binary fraction 
were changed}.  This is also true of the magnitude limit and stellar age, which 
effectively change the mean of the log-normal and therefore behave similarly to 
$\mu_{\log P}$. We therefore conclude that the parameters $\mu_{\log P}$, 
$\sigma_{\log P}$, magnitude limit $M_{lim}$ and stellar age $t_g$ are all 
nearly degenerate with binary fraction over the scale of km/s.

The degeneracy of the period distribution parameters with binary fraction also 
holds for the two-epoch velocity distribution $g_b(\log|\Delta v|;\Delta t)$, 
since this also has a log-normal form for km/s velocities. By 
eq.~\ref{eq:F_dv}, therefore, the same degeneracy holds for the threshold 
fraction $F$.  The effect of this degeneracy on the threshold fraction and its 
implications for constraining the binary distribution of periods will be 
explored in further detail in section \ref{sec:period_dist}.

In this section we will consider the threshold fraction $F_0$ with a fixed time 
interval of 1 year and without measurement error, i.e.  $F_0 = F(\Delta 
v|\Delta t=$1 year$,\sigma_{2e}=0)$ (section \ref{sec:threshold_fraction}).  
There is no loss of generality in this; as we demonstrated in section 
\ref{sec:multi_epoch}, the threshold fraction for any given time interval 
$\Delta t$ can be estimated by a likelihood analysis even if measurements are 
taken at various different epochs and with various different measurement 
errors.  However, if the threshold fraction $F$ is measured directly for a 
fixed time interval, it is necessary to account for the effect of measurement 
error on $F$; we will address this later in the section.

In the absence of measurement error, by definition $F_0$ scales linearly with 
the binary fraction $B$ (eq.~\ref{eq:F_dv_noerr}). Furthermore, because of the 
near-degeneracy of the period distribution parameters with binary fraction, 
$F_0$ also scales linearly with $\mu_{\log P}$ and $\sigma_{\log P}$ to good 
approximation over their viable range of values (roughly 1-3 with $P$ in years; 
see section \ref{sec:period_dist}).  The essential point is that a similar 
relationship holds for the velocity dispersion if a high-velocity cutoff is 
used, e.g.  at $v_c = 3\sigma$, since the degeneracy approximately holds for 
velocities $v < v_c$.  It follows that if velocity outlier stars are excluded 
in determining velocity dispersion, the extra dispersion due to binaries can be 
determined from the threshold fraction $F_0$ with reasonable confidence even if 
the parameters $B$, $\mu_{\log P}$, and $\sigma_{\log P}$ are entirely unknown.

We demonstrate this by simulating galaxies with various intrinsic dispersions 
and characterized by different binary populations. The dispersion $\sigma$ is 
calculated by iteratively discarding stars with velocities larger than 
3$\sigma$; on the first iteration the biweight is used to estimate the 
dispersion, and the dispersion is then calculated on every subsequent iteration 
until all the remaining stars have velocities that lie within $3\sigma$. In 
order to make the statistical error negligible, we used a very large ``sample'' 
of 100,000 stars.  We also calculate the threshold fraction $F_0$ for the same 
data set, for which we picked a threshold $\Delta v = 5$ km/s and time interval 
$\Delta t = 1$ year.

\begin{figure}
	\includegraphics[height=1.0\hsize,angle=-90]{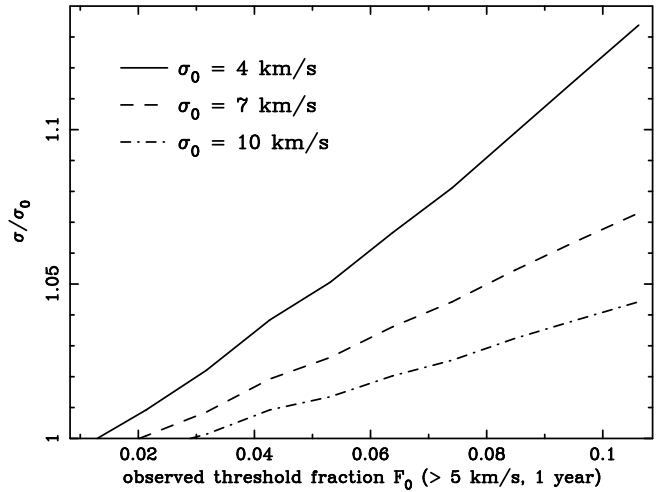}
\caption{Ratio of measured velocity dispersion $\sigma$ over the intrinsic 
dispersion $\sigma_0$, plotted with respect to threshold fraction $F_0$ for 
different binary fractions in galaxies of intrinsic dispersions $\sigma_0 = 5$, 
7, and 10 km/s.  The measured dispersions were calculated by an interative 
$3\sigma$-clipping routine, and the threshold fraction $F_0$ denotes the 
fraction of stars with observed change in velocity greater than a threshold 
$\Delta v = 5$ km/s after a time $\Delta t = 1$ year between measurements, 
assuming zero measurement error. The fiducial period distribution ($\mu_{\log 
P}=2.23$, $\sigma_{\log P}=2.3$, $P$ in years) is assumed, and the binary 
fraction is varied between 0.1 and 1.  }
\label{fig:fdisp_sigreal_b}
\end{figure}

First, we assume the fiducial binary period distribution ($\mu_{\log P}$ and 
$\sigma_{\log P}$) and vary the binary fraction from $B=0.1$ to 1. In 
fig.~\ref{fig:fdisp_sigreal_b} we plot the ratio $\sigma/\sigma_0$ of measured 
dispersion over the intrinsic dispersion as a function of threshold fraction, 
for galaxies with intrinsic dispersions of 4, 7, and 10 km/s. We used an 
absolute magnitude limit $M_{lim} = 3$, however the graph remains virtually 
unchanged for other magnitude limits because of the near-degeneracy of 
magnitude limit with binary fraction discussed at the beginning of this 
section.  We see that for a given intrinsic dispersion, the observed threshold 
fraction can be mapped in a one-to-one way to the intrinsic dispersion. The 
relation shown in the graph also holds regardless of the age of the stellar 
population, again because of the degeneracy of age with binary fraction.  

\begin{figure}
	\includegraphics[height=1.0\hsize,angle=-90]{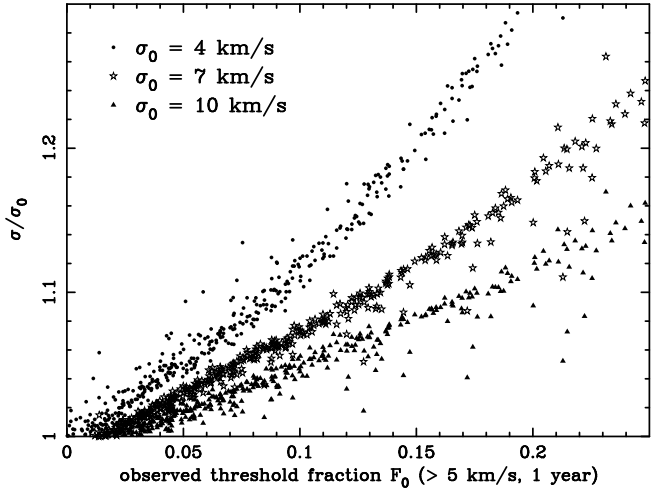}
\caption{Ratio of measured velocity dispersion $\sigma$ over the intrinsic 
dispersion $\sigma_0$, plotted with respect to threshold fraction $F_0$ for 
different binary populations in galaxies of intrinsic dispersions $\sigma_0 = 
5$, 7, and 10 km/s.  The measured dispersions were calculated by an interative 
$3\sigma$-clipping routine, and the threshold fraction $F_0$ denotes the 
fraction of stars with observed change in velocity greater than a threshold 
$\Delta v = 5$ km/s after a time $\Delta t = 1$ year between measurements, 
assuming zero measurement error. Each point represents a different binary 
population with its own binary fraction and period distribution; we plotted the 
points over a grid of values, with binary fraction $B$ ranging from 0.2 to 1, 
$\mu_{\log P}$ from -1 to 4 (in log($P$/year)) and $\sigma_{\log P}$ from 0.5 
to 4. We show at the end of section \ref{sec:correcting_dispersion} how $F_0$ 
can be inferred from observations accounting for measurement errors. }
\label{fig:fdisp_sigreal}
\end{figure}

Next we repeat the procedure over a grid of values for the parameters $B$, 
$\mu_{\log P}$, and $\sigma_{\log P}$, and for each point we plot the ratio 
$\sigma/\sigma_0$ with respect to the threshold fraction $F_0$. The results are 
plotted in fig.~\ref{fig:fdisp_sigreal}, again for galaxies with intrinsic 
dispersions of 4, 7, and 10 km/s.  We see that for each group, the graph forms 
a tight relation for all but the most extreme values of the period distribution 
parameters. In plotting these points we varied $B$ from 0.2 to 1, $\mu_{\log 
P}$ from -1 to 4, and $\sigma_{\log P}$ from 0.5 to 4 (with $P$ in years).  The 
lowermost points of each group are the points for with $\sigma_{\log P}$ has 
its smallest value of 0.5, producing only a very small number of short-period 
binaries. The uppermost points are the points for which $\mu_{\log P}$ has its 
smallest value, so the period distribution is shifted toward short periods. For 
these extreme values, the velocity distribution becomes distorted into a 
distinctly non-Gaussian shape so these can be considered highly improbable 
configurations. We have also varied the ellipticity distribution parameters 
$\bar e$, $\sigma_e$ (section \ref{sec:derivations}) and find that the tight 
correlation in fig.~\ref{fig:fdisp_sigreal} is unchanged, and although the 
amount of scatter increases slightly, the correction still holds to within a 
few percent accuracy.

We find that the points plotted in fig.~\ref{fig:fdisp_sigreal} are well fit by 
a line plus an exponential function, provided that outlier points are 
discarded.  To define ``outliers'', first we divide the domain $F_0\in [0,0.2]$ 
into sections small enough so the graph is approximately linear within each 
section.  We then further divide each section into two subsections and 
calculate the median and median absolute deviation (MAD) of the $y$-values of 
the points in each subsection. Next we draw lines through the two points 
defined by the median $\pm$ twice the MAD of each of the two subsections, 
taking the center of the subsection as their $x$-value.  The plotted points 
that lie outside the region defined by these lines represent extreme and highly 
improbable period distributions, and are therefore discarded. We fit the 
remaining points and repeat the procedure for galaxies of dispersions ranging 
from 3-12 km/s.

\begin{figure}
	\includegraphics[height=1.0\hsize,angle=-90]{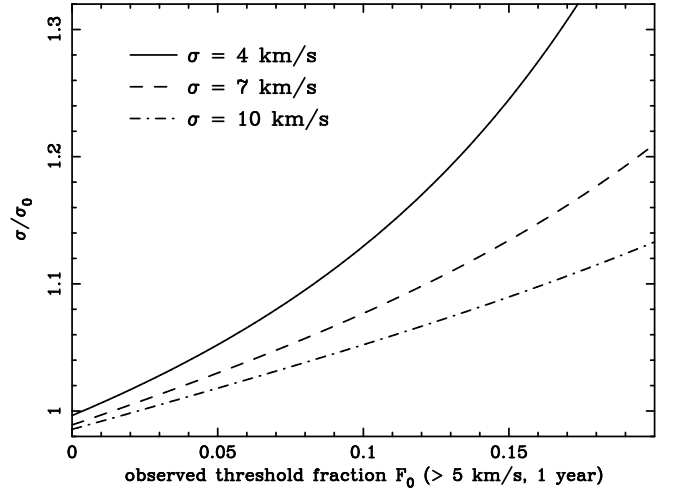}
\caption{Ratio of measured velocity dispersion $\sigma$ over the intrinsic 
dispersion $\sigma_0$, plotted with respect to threshold fraction $F_0$ for 
\emph{fixed measured dispersions} $\sigma$. These curves were found by fitting 
graphs like those shown in fig.~\ref{fig:fdisp_sigreal}, then transforming to 
fixed values of the measured dispersion $\sigma$. The measured dispersions in 
fig.~\ref{fig:fdisp_sigreal} were calculated by an interative 
$3\sigma$-clipping routine, and the threshold fraction $F_0$ denotes the 
fraction of stars with change in velocity greater than a threshold $\Delta v = 
5$ km/s after a time $\Delta t = 1$ year between measurements, with zero 
measurement error. We show at the end of section 
\ref{sec:correcting_dispersion} how $F_0$ can be inferred from observations 
accounting for measurement errors.  }
\label{fig:fdisp_sigm}
\end{figure}

The plots in fig.~\ref{fig:fdisp_sigreal} are not directly applicable to real 
data because each graph was plotted for a fixed intrinsic dispersion 
$\sigma_0$, which is unknown (and is in fact what we are attempting to 
calculate!).  We therefore use our fits together with a root-finding procedure, 
interpolating the fitting parameters in $\sigma_0$, to draw similar graphs at 
fixed values of $\sigma$. A few resulting curves are plotted in 
fig.~\ref{fig:fdisp_sigm}.  Again, we find these curves are well fit by a line 
plus exponential,

\begin{equation}
\label{eq:sigratio_fitting_function}
\frac{\sigma}{\sigma_0} = a(\sigma) + b(\sigma)F_0 + c(\sigma)\left[e^{F_0/0.1}-1\right]
\end{equation}
where $F_0$ is the threshold fraction at 5 km/s.  We also find fitting 
functions for the parameters $a(\sigma)$, $b(\sigma)$, and $c(\sigma)$ which 
fit well for dispersions $\sigma$ ranging from 4 km/s to 10 km/s.  Defining 
$\Delta\sigma = \sigma - 4$ km/s, we find:

\begin{eqnarray}
\label{eq:sigratio_fit_parameter_a}
a(\sigma) & = & 0.988e^{-0.0007\Delta\sigma} \\
\label{eq:sigratio_fit_parameter_b}
b(\sigma) & = & 0.576 - 0.08\Delta\sigma + 0.772\left(1-e^{-0.1\Delta\sigma}\right) \\
\label{eq:sigratio_fit_parameter_c}
c(\sigma) & = & 0.043e^{-0.247\Delta\sigma}
\end{eqnarray}

These formulas hold for any magnitude limit and stellar age, and the threshold 
fraction $F_0$ refers here to a velocity threshold of 5 km/s, time interval of 
1 year, and zero measurement error.

How are these formulas adjusted if a different velocity threshold is desired?  
Ideally, one should use the smallest possible threshold that is not 
significantly affected by measurement error--this will include the most stars 
and therefore have a smaller scatter compared to higher thresholds. To use a 
different threshold $\Delta v$, the value of $F_0$ in 
eq.~\ref{eq:sigratio_fitting_function} must be scaled by the ratio $F_0(\Delta 
v)/F_0(5~\textrm{km/s})$.  This ratio can be calculated by using Monte Carlo 
realizations to plot the threshold fraction as a function of threshold, which 
we will do in section \ref{sec:multi_epoch} (figs.~\ref{fig:dvchist}, 
\ref{fig:dvchist_dt}).  Again, the degeneracy of magnitude limit and period 
distribution parameters with binary fraction ensures that this ratio is 
virtually independent of the model parameters and magnitude limit, provided one 
does not transform to thresholds that are too high ($>$ 10 km/s).  We find that 
for thresholds in the range 4 km/s $< \Delta v < $ 10 km/s, the ratio 
$F_0(\Delta v)/F_0(5~\textrm{km/s})$ can be fit by the function

\begin{equation}
\label{eq:threshold_scaling}
\frac{F_0(\Delta v)}{F_0(5~\textrm{km/s})} = a + b e^{-\Delta v/\Delta v_s}
\end{equation}
where the best-fit parameters are $a = 0.0725$, $b = 1.897$, and $\Delta v_s = 
6.947$ km/s. Thus to find the correction to the dispersion in terms of a given 
velocity threshold $\Delta v$, one substitutes eq.~\ref{eq:threshold_scaling} 
into eq.~\ref{eq:sigratio_fitting_function} so that the fit is in terms of 
$F_0(\Delta v)$.

%Similarly, to estimate the error in the intrinsic dispersion one substitutes 
%eq.~\ref{eq:threshold_scaling} into eq.~\ref{eq:sigma_intrinsic_error} for 
%both $F_0$ and $\delta F$. The Poisson error $\delta F \approx 
%\sqrt{F_0(\Delta v)/N}$ should be used, then propagated to $\delta F$(5 km/s) 
%by using eq.~\ref{eq:threshold_scaling}.

The sampling error in the intrinsic dispersion $\sigma_0$ determined by this 
procedure can be estimated by noting that for a two-epoch sample of $N_{2e}$ 
stars, the Poisson error in the threshold fraction is $\delta F \approx 
\sqrt{F_0(\Delta v)/N_{2e}}$. Let us assume the dispersion is measured in a 
larger single-epoch sample of $N$ stars; then we can make the approximation 
that the sampling errors in $\sigma$ and $F$ are weakly correlated so they add 
in quadrature. Propagating the error using 
eq.~\ref{eq:sigratio_fitting_function} gives the sampling error in the 
intrinsic dispersion,

\begin{equation}
\label{eq:sigma_intrinsic_error}
\left(\frac{\delta \sigma_0}{\sigma_0}\right)^2 \approx \frac{\left(\sigma/\sigma_0\right)^2}{2N} + 
\left|\frac{\sigma_0}{\sigma}\frac{\partial\left(\sigma/\sigma_0\right)}{\partial F_0}\right|^2 
\frac{F_0(\Delta v)}{N_{2e}}
\end{equation}
%\begin{equation}
%\label{eq:sigma_intrinsic_error}
%\left(\frac{\delta \sigma_0}{\sigma_0}\right)^2 \approx \frac{\Sigma^2}{2N} + 
%\left|\frac{1}{\Sigma}\frac{\partial\Sigma}{\partial F_0}\right|^2 
%\frac{F_0(\Delta v)}{N_{2e}}
%\end{equation}
where the second term in eq.~\ref{eq:sigma_intrinsic_error} is the two-epoch 
sampling error. For example, if the single-epoch sample contains $N=500$ stars 
and the measured threshold fraction is $F_0 = 0.1$ in a two-epoch subset of 
$N_{2e}=100$ stars, the intrinsic dispersion $\sigma_0$ can be determined to 
within $\approx $ 5\% for dispersions $\sigma > 4$ km/s.  
Eq.~\ref{eq:sigma_intrinsic_error} was tested with Monte Carlo simulations and 
found to be accurate to within $\approx$ 0.8\% for $N = N_{2e}$; in the above 
example, the formula is accurate to within $\approx$ 0.5\%. However, the 
fractional error in $\sigma_0$ cannot be made smaller than $\approx$ 0.5\% 
because of the inherent uncertainty in the binary population model represented 
by the width of the plots in fig.~\ref{fig:fdisp_sigreal}. We find that for an 
observed threshold fraction $F = 0.1$, the two-epoch sampling error is smaller 
than the single-epoch error unless $N/N_{2e} > 5$.  Given a measured threshold 
fraction $F(\Delta v)$, the two-epoch sampling error is larger relative to the 
single-epoch error for smaller measured dispersions.

Finally, we address the issue of measurement error. As mentioned above, the 
best approach is to estimate the error-free threshold fraction $F_0$ by a 
Bayesian or maximum-likelihood approach (see section \ref{sec:multi_epoch}), in 
which case measurement error need not be considered here. However, if $F$ is 
calculated directly from the data, then measurement error must be considered in 
the above formulas. We find that given a 2-epoch measurement error 
$\sigma_{2e}$ (eq.~\ref{eq:sigma_2m}), the threshold fraction $F(\Delta 
v|\sigma_{2e})$ is related to the threshold fraction without measurement error 
$F_0(\Delta v)$ by a linear transformation---yet another consequence of the 
degeneracy between binary fraction and the period distribution parameters (see 
appendix \ref{sec:appendix_a} for a derivation of this result).  The 
transformation takes the form:

\begin{equation}
\label{eq:F_measurement_error_key_eq}
F(\Delta v|\sigma_{2e}) \approx \textrm{erfc}\left[\frac{\Delta 
v}{\sqrt{2}\sigma_{2e}}\right] + \beta\left(\frac{\Delta v}{\sigma_{2e}}\right) 
F_0(\Delta v)
\end{equation}

This result (together with eq.~\ref{eq:threshold_scaling}) has been tested by 
using the Monte Carlo simulation to produce plots similar to 
fig.~\ref{fig:fdisp_sigreal} for different velocity thresholds and measurement 
errors. The approximate analytic form of $\beta$ can also be derived (see 
appendix \ref{sec:appendix_a}); using this together with the Monte Carlo plots 
to map $\beta$, we find that $\beta$ is well fit by the following function 
(appendix \ref{sec:appendix_a}):

\begin{equation}
\label{eq:beta_fitting_function}
\beta(x) = \left(1 + a e^{-\frac{x}{x_s}}\right) \left\{1 - \bar \kappa \cdot
\textrm{erfc}\left(\frac{x}{\sqrt{2}}\right)\right\}
\end{equation}
where $x = \Delta v/\sigma_{2e}$ and the best-fit parameters are $a = 0.05$, 
$x_s = 5$, and $\bar \kappa = 1.3$. Thus to find the correction to the 
dispersion with a given measurement error, one substitutes the error-free 
threshold fraction $F_0$ in terms of $F(\Delta v|\sigma_{2e})$ (given by 
eq.~\ref{eq:F_measurement_error_key_eq}) into 
eq.~\ref{eq:sigratio_fitting_function}. The effect of the measurement error on 
the dispersion must also be taken into account by making the substitution 
$\sigma^2 = \sigma_{meas}^2 - \sigma_m^2$ into eq.  
\ref{eq:sigratio_fitting_function}, where $\sigma_{meas}$ is the measured 
dispersion and $\sigma_m$ is the measurement error. 

In summary, the velocity dispersion of a dwarf spheroidal sample can be 
corrected for binaries by the following method:

1. Measure the threshold fraction $F$ for a particular threshold velocity and a 
time interval $\Delta t$ = 1 year.  This can be done in two ways: the threshold 
fraction can be measured directly, in which case one should use the smallest 
possible threshold that is not unduly affected by measurement error; this is 
approximately $\Delta v \approx 2 \bar \sigma_{2e}$ where $\bar\sigma_{2e} = 
\sqrt{2}\bar\sigma_m$ is the median two-epoch measurement error. An alternative 
(and more rigorous) approach is to estimate the measurement error-free 
threshold fraction $F_0$ by a likelihood analysis.  This procedure is 
demonstrated in section \ref{sec:multi_epoch}.

2. Measure the velocity dispersion $\sigma_{meas}$ of the sample by an 
iterative 3$\sigma$-clipping routine. Correct the dispersion for measurement 
error to find the error-free dispersion $\sigma$.

3. If the chosen velocity threshold is different from 5 km/s, scale the 
threshold fraction $F_0$(5 km/s) in eq.~\ref{eq:sigratio_fitting_function} 
according to eq.~\ref{eq:threshold_scaling}.

4. If the threshold fraction is measured directly, one must use 
eq.~\ref{eq:F_measurement_error_key_eq} to relate the error-free threshold 
fraction $F_0(\Delta v)$ in eq.~\ref{eq:sigratio_fitting_function} to the 
threshold fraction with measurement error, $F(\Delta v|\sigma_{2e})$.

5. After substituting $F_0$(5 km/s) in terms of $F(\Delta v|\sigma_{2e})$ 
(given by steps 3 and 4 above) into eq.~\ref{eq:sigratio_fitting_function}, use 
eq.~\ref{eq:sigratio_fitting_function} together with 
eqs.~\ref{eq:sigratio_fit_parameter_a}-\ref{eq:sigratio_fit_parameter_c} to 
find the intrinsic velocity dispersion $\sigma_0$. The sampling error in 
$\sigma_0$ can be estimated from eq.~\ref{eq:sigma_intrinsic_error}.

Finally, from fig.~\ref{fig:fdisp_sigreal} we can estimate an approximate upper 
bound to the dispersion introduced by binaries in dwarf spheroidals.  First we 
note that given an absolute magnitude limit $M_{lim}<1$ and a stellar 
population older than 1 Gyr, and if the distribution of binary orbital 
parameters mirrors that of the solar neighborhood, the threshold fraction $F_0$ 
cannot be larger than 0.12 even if the binary fraction is 1 
(fig.~\ref{fig:dvchist_age}). By analyzing multi-epoch data in the Fornax, 
Carina, Sculptor, and Sextans dwarf spheroidals
(\citealt{walker11-09}) we find they all have threshold fractions smaller than 
0.15, and only Fornax has $F \gtrsim 0.1$.  Assuming this is generally the case 
even for ultra-faint dwarf spheroidals with intrinsic dispersions greater than 
4 km/s, we conclude from fig.~\ref{fig:fdisp_sigreal} that the measured 
velocity dispersions of these galaxies are unlikely to be inflated by more than 
20\%. However, even if the dispersion of a particular galaxy is inflated by 
more than 20\%, the correction due to binaries can be readily discerned by 
making repeat measurements and applying the method outlined above.

\section{Bayesian analysis of single-epoch velocity data}\label{sec:bayesian}

\begin{figure*}
	\centering
	\subfigure[$\sigma_{real}$ = 4 km/s, 100 stars]
	{
		\includegraphics[height=0.32\hsize,width=0.33\hsize,angle=-90]{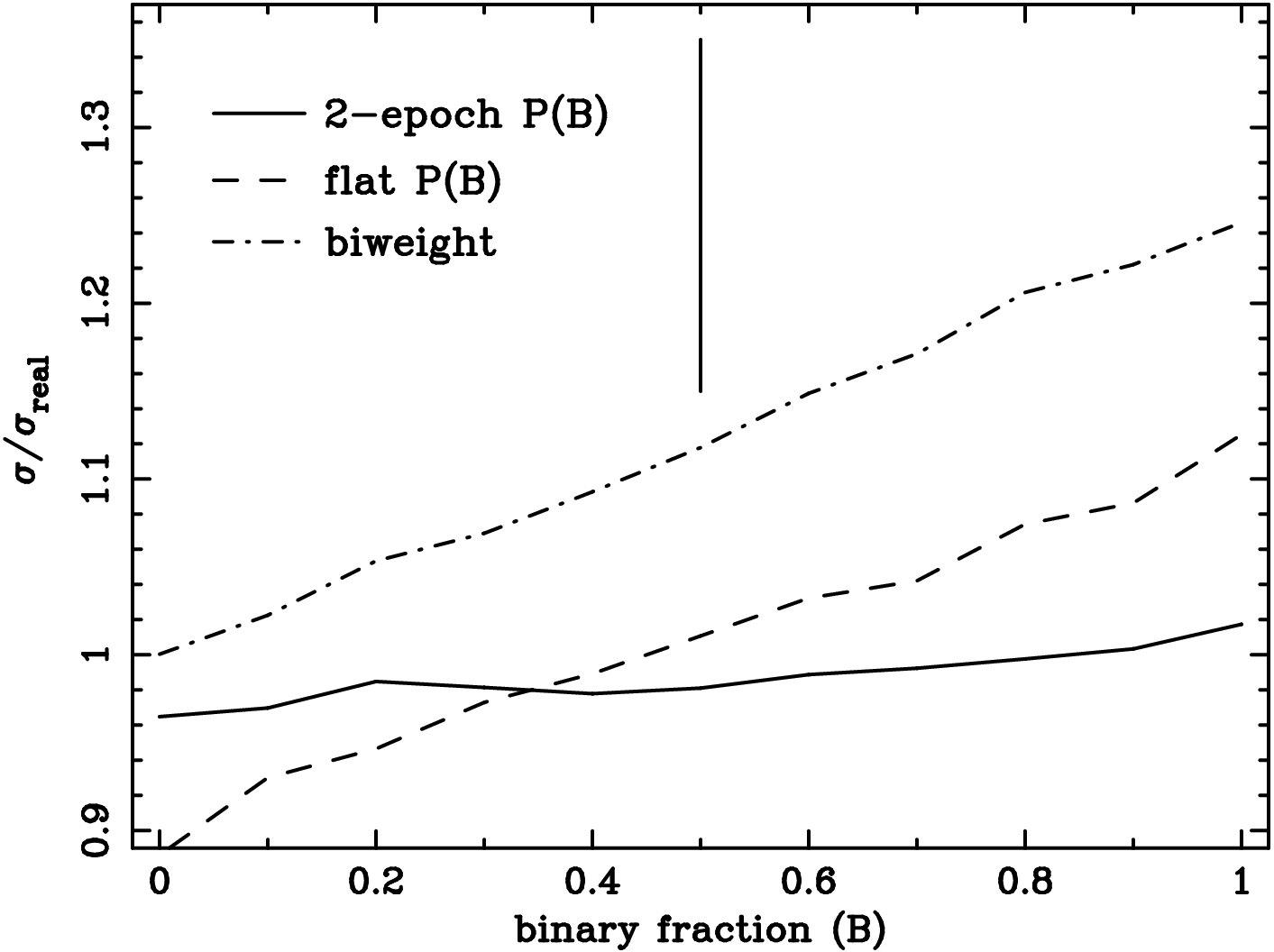}
		\label{fig:4kms_100stars}
	}
	\subfigure[$\sigma_{real}$ = 4 km/s, 500 stars]
	{
		\includegraphics[height=0.32\hsize,width=0.33\hsize,angle=-90]{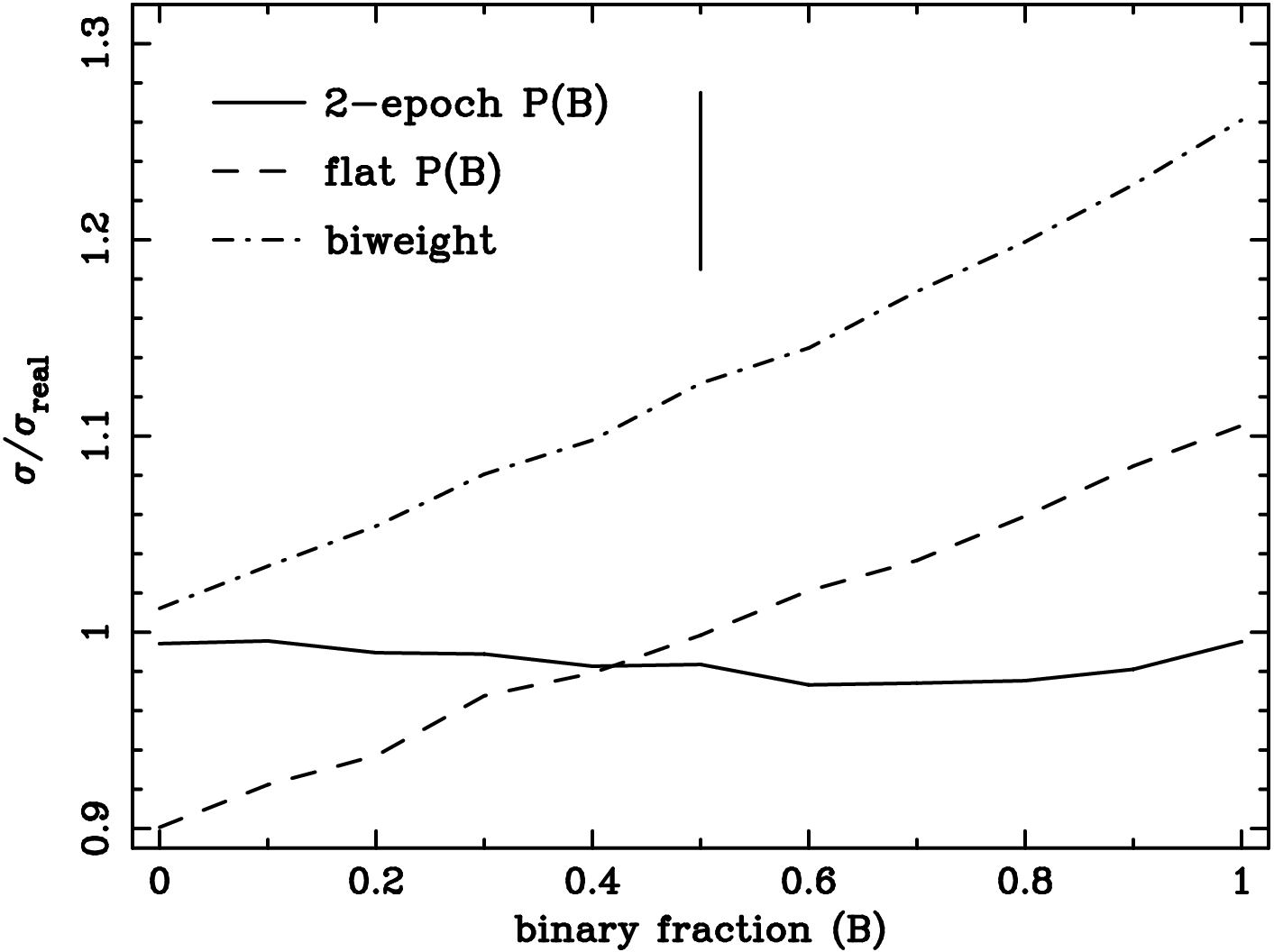}
		\label{fig:4kms_500stars}
	}
	\caption{Best-fit values for the dispersion $\sigma_{fit}$ in a simulated 
galaxy with dispersion $\sigma_{real}=4$ km/s and different binary fractions 
$B$.  The solid vertical line gives the error bar, equal to one standard 
deviation in $\sigma_{fit}$ values calculated in 300 random realizations, which 
is similar for all points on the graph.  Solid line uses a uniform prior in 
$B$, while dashed line uses a prior $P(B)$ obtained from multi-epoch 
observations of the stars.  Dot-dashed line is calculated from the biweight 
robust estimator.  We adopted a measurement error of 2 km/s for all stars.}

	\label{fig:4kms_sigfits}
\end{figure*}

\begin{figure*}
	\centering
	\subfigure[$\sigma_{real}$ = 10 km/s, 100 stars]
	{
		\includegraphics[height=0.32\hsize,width=0.33\hsize,angle=-90]{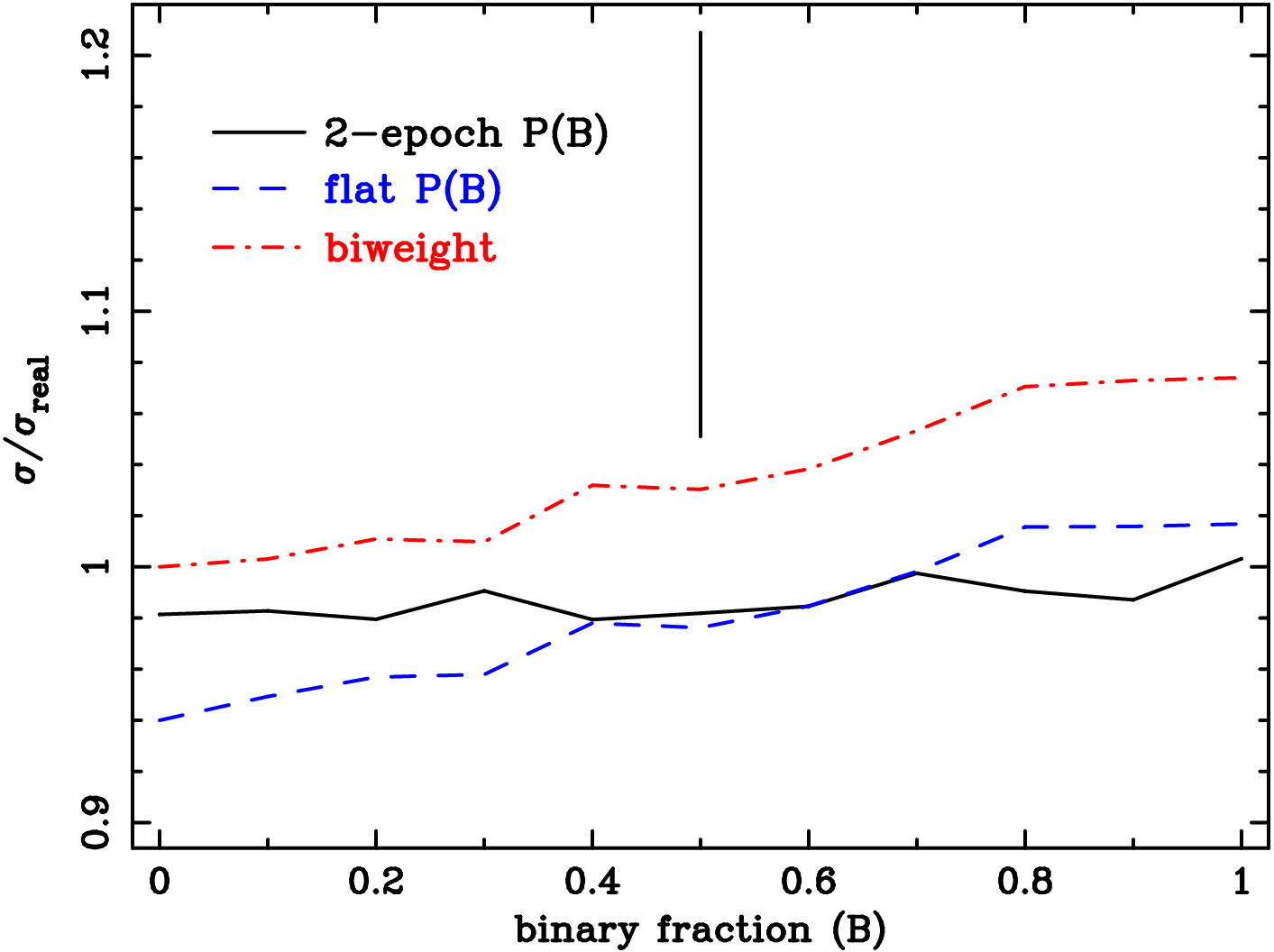}
		\label{fig:10kms_100stars}
	}
	\subfigure[$\sigma_{real}$ = 10 km/s, 500 stars]
	{
		\includegraphics[height=0.32\hsize,width=0.33\hsize,angle=-90]{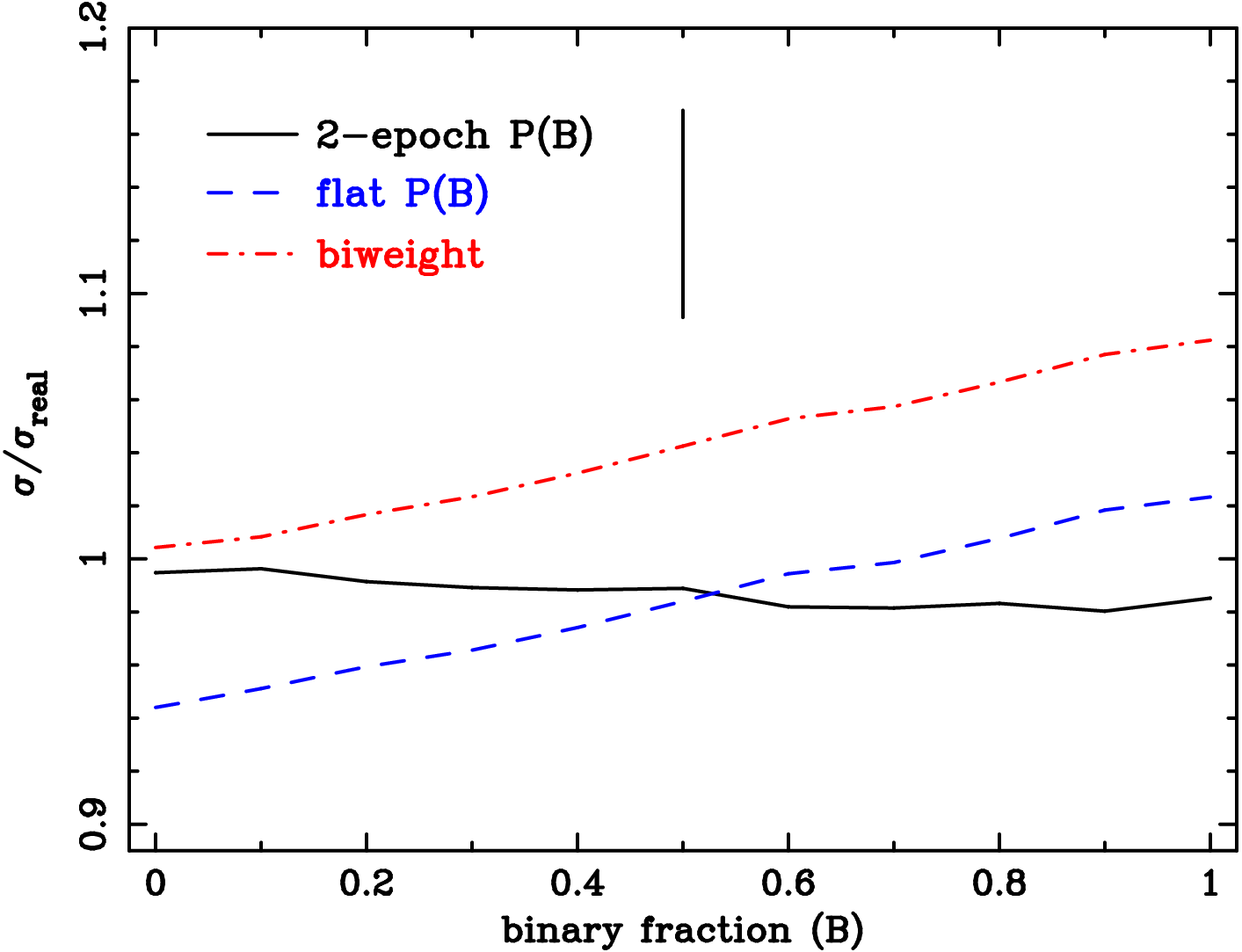}
		\label{fig:10kms_500stars}
	}
	\caption{Best-fit values for the dispersion $\sigma_{fit}$ in a simulated 
galaxy with dispersion $\sigma_{real}=10$ km/s and different binary fractions 
$B$.  Error bar gives one standard deviation in $\sigma_{fit}$ values 
calculated in 300 random realizations, which is similar for all points on the 
graph.  Solid line uses a uniform prior in $B$, while dashed line uses a prior 
$P(B)$ obtained from multi-epoch observations of the stars.  Dot-dashed line is 
calculated from the biweight robust estimator.  We adopted a measurement error 
of 2 km/s for all stars.}
	\label{fig:10kms_sigfits}
\end{figure*}

In this section we discuss the problem of constraining properties of a binary 
population with radial velocity measurements taken at a single epoch. We 
approach this by fitting the likelihood for binary stars (eq.~\ref{f_v}) to a 
simulated data sample generated from a Monte Carlo simulation for galaxies with 
dispersions 4 km/s and 10 km/s.  Both galaxies were given a binary fraction $B$ 
= 0.5 and period distribution parameters equal to their fiducial values 
$\mu_{\log P}$ = 2.23, $\sigma_{\log P}$ = 2.3 ($P$ in years). The simulated 
velocities were generated with a measurement error of 2 km/s.  Using 
eq.~\ref{f_v} together with Bayes' Theorem, marginal posterior distributions in 
the dispersion $\sigma$ and binary fraction $B$ can be obtained.  The maxima of 
these distributions are taken as the best-fit values $\sigma_{fit}$ and 
$B_{fit}$.  

In the absence of any constraint on the binary fraction from multi-epoch data, 
we assume a uniform prior in the binary fraction.  However, if repeat 
measurements do exist for some subset of the data sample, these can be used to 
better constrain the binary fraction. The most rigorous approach would use a 
joint likelihood $L(v_1,\Delta v|\Delta t)$, which can be generated from the 
Monte Carlo. This would have the advantage that individual binary stars which 
are inflating the dispersion \emph{and} exhibit a large velocity change $\Delta 
v$ would be recognized as such, and weighted accordingly. While this method may 
be necessary for obtaining constraints in small data sets of less than 100 
stars, for larger samples we can adopt a simpler approach: first derive a 
posterior $P(B)$ in binary fraction by the multi-epoch analysis outlined in 
section \ref{sec:multi_epoch}, and subsequently take $P(B)$ as a \emph{prior} 
in $B$ for the single-epoch analysis. The usefulness of this method depends 
critically on the size of the multi-epoch sample, as this determines the 
constraint on binary fraction (see fig.~\ref{fig:approx}).

To evaluate this method, it is important to know how the best-fit dispersion 
$\sigma_{fit}$ obtained by this procedure may differ from the intrinsic 
dispersion $\sigma_{real}$ due to small number statistics.  To investigate 
this, we analyze samples consisting of 100 and 500 stars and repeat the 
procedure over 300 randomly generated realizations of each sample.  The range 
of $\sigma_{fit}$ values obtained for galaxies with dispersions of 4 km/s and 
10 km/s are plotted in fig.~\ref{fig:4kms_sigfits} and 
fig.~\ref{fig:10kms_sigfits}, respectively.  Also included are the results if 
the prior in $B$ is taken from multi-epoch observations of the same stars in 
the sample, in the manner outlined above.  For comparison, we also plot the 
dispersion obtained from the biweight robust estimator, which is roughly 
comparable to the dispersion obtained by using a using a 3$\sigma$-clipping 
routine.  This plot shows the biweight is biased to large dispersions by an 
amount which depends on the binary fraction; furthermore, the bias for a 4 km/s 
dispersion is much larger compared to the error bars than for a 10 km/s 
dispersion. By comparison, the best-fit dispersion $\sigma_{fit}$ using a 
uniform prior in $B$ is biased by a smaller amount which depends on the true 
binary fraction, up to 10\%.  However, if the prior $P(B)$ is calculated from 
multi-epoch samples of as few as 100 stars, the bias is almost entirely 
removed. Thus, even a fairly weak constraint on the binary fraction is 
sufficient to remove most of the bias.

Unfortunately, the likelihood analysis outlined above is of limited utility in 
actual data sets because of contamination by foreground Milky Way stars.
The usual criteria to determine membership of a star include its location on 
the color-magnitude diagram, metallicity, and radial velocity.  Outliers in the 
distribution of these variables are either excluded, or else weighted by a low 
membership probability assigned according to a specific algorithm (cf.  
\citealt{walker11-09}). However, among the stars with single-epoch 
measurements, only those stars which lie on the high-velocity tail of the 
velocity distribution can help constrain the binary fraction. If even a small 
number of high-velocity binary stars are excluded from the data or weighted by 
a low membership probability, the inferred binary fraction will be biased to 
low values.

In principle this problem can be resolved by including the velocity 
distribution of foreground Milky Way stars in the likelihood analysis, obtained 
by a kinematic model of the disk and bulge components (\citealt{robin2003}).  
The success of this method will depend critically on the degree of overlap 
between the two distributions, which is partly determined by the systemic 
velocity of the background galaxy.  It is also essential that binarity in the 
Milky Way is accounted for, since this adds a substantial high-velocity tail to 
the foreground velocity distribution.
%This can be accomplished by averaging the Milky Way velocity distribution over 
%the distribution of binary velocities in a manner similar to 
%eq.~\ref{fv_binary_part}, with the additional complication that magnitude cuts 
%at different distances must be taken into account and the result must be 
%integrated along the line-of-sight. The contribution of binaries will be more 
%substantial than in the background galaxy because it is dominated by younger 
%main sequence stars.  Calculating a Milky Way likelihood in this manner does, 
%however, have the advantage that the binary population is well constrained in 
%the solar neighborhood (\citealt{duquennoy1991}).  The total likelihood would 
%take the form
%
%\begin{equation}
%L(v|f,\sigma_0,B_{mw},B) = (1-f) L_{mw}(v|B_{mw}) + f L(v|B,\sigma_0)
%\end{equation}
%
%where $f$ is the total fraction of member stars and $B_{mw}$ is the Milky Way 
%binary fraction. This likelihood can be readily extended into a joint 
%likelihood in velocity and metallicity, which can help in distinguishing member 
%from non-member stars.
Accounting for binarity in the likelihoods of both the foreground 
and background stars may also lead to improved membership probabilities when 
combined with the expectation maximation algorithm of \cite{walker11-09}.  
Including both binarity and foreground Milky Way stars in a likelihood analysis 
is certainly of interest for obtaining the best possible constraints from 
single-epoch velocity data, but is beyond the scope of the present work.

\section{Fitting function for binary likelihood}\label{sec:fitting_function}

For the purpose of making analytic calculations as well as doing a likelihood 
analysis, it is useful to have a fitting function for the binary velocity 
distribution. This can be found by investigating the exponent of the binary 
velocity distribution. We find that the distribution indeed follows a 
log-normal but the dispersion changes at a velocity scale $v_0$ due to the 
suppression of binaries by Roche lobe overflow.  This suggests the following 
fitting function for the binary distribution of velocities:

\begin{eqnarray}
\label{eq:flogv_fitting_function}
\lefteqn{ \nonumber
f_b(\log |v|) ~ = ~ (1-\mathcal{N})\mathcal{D}(\log|v|) } \\
&& + ~ \mathcal{N} \exp\left\{\frac{-1}{2\sigma_{\log P}^2} \left[\mu_{\log P} + 3\mathcal{G}(\log|v|)\right]^2\right\},
\end{eqnarray}

As we discuss below, all the fitting parameters are contained in the function $\mathcal{G}(\log|v|)$:

\begin{equation}
\mathcal{G}(\log|v|) = 
\left\{\begin{array}{ll}
 \alpha + s \log |v|, \quad & |v| \leq v_0 \\
 \alpha + s'\log\left|\frac{v}{v_0}\right| + s\log v_0, \quad & |v| > v_0.
\end{array} \right.
\label{eq:g_fitting_function}
\end{equation}

For velocities $|v| < v_0$, the log-normal dispersion is given by 
$\sigma_{\log|v|} = \frac{\sigma_{\log P}}{3s}$. The destruction of binaries 
due to Roche-lobe overflow becomes important for velocities $|v| > v_0$, for 
which the log-normal dispersion steepens to $\sigma_{\log|v|} = 
\frac{\sigma_{\log P}}{3s'}$.

The normalizing factor $\mathcal{N}$ (given by eq.~\ref{normalizing_factor} 
below) is not a fitting parameter, but rather is determined by the other 
fitting parameters along with $\mu_{\log P}$ and $\sigma_{\log P}$.  The 
function $\mathcal{D}$, whose parameters are fixed, is given by

\begin{equation}
\label{delta_fn}
\mathcal{D}(\log|v|) = \frac{e^{-(\log|v| - \epsilon)^2/2\sigma_\delta^2}}{\sqrt{2\pi\sigma_\delta^2}},
\end{equation}
and is effectively a smoothed $\delta$-function at very small $v$.  With $v$ in 
units of AU/year, we find $\epsilon = -3$ and $\sigma_\delta = 1$ works well 
with only slight deformation of the distribution at small $v$, however the 
exact values of these parameters are unimportant so long as the width and mean 
of $\mathcal{D}$ is sufficiently small. Although stars at such small velocities 
give a negligible contribution to the dispersion,  including the smoothed 
$\delta$-function is crucial as it gives the appropriate weight to the 
small-velocity stars.

The normalizing factor $\mathcal{N}$ in eq.~\ref{eq:flogv_fitting_function} is 
given by

\begin{equation}
\label{normalizing_factor}
\mathcal{N} = \frac{3}{\sqrt{2\pi\sigma_{\log P}^2}}\times \frac{0.8}{\eta(\mu_{\log P},\sigma_{\log P})},
\end{equation}

\begin{equation}
\eta(\mu_{\log P},\sigma_{\log P}) = \frac{1}{2s}\textrm{erfc}[-x(s)] + \frac{1}{2s'}\textrm{erfc}[x(s')],
\end{equation}

\begin{equation}
\label{normalizing_x_factor}
x(s) = \frac{3s\log|v_0| + 3\alpha + \mu_{\log P}}{\sqrt{2}\sigma_{\log P}}.
\end{equation}

\begin{table}[ht]
 \begin{tabular}{ccccccc}
  \hline
	$M_V$ & $\alpha$ & $s$ & $s'$ & $v_0$ & $\mathcal{N}$ & $\sigma_b$ \\
	\hline
	-1 & 0.260 & 0.998 & 2.408 & 2.500 & 0.785 & 2.741 \\
	0 & 0.210 & 0.888 & 1.920 & 2.445 & 0.878& 3.460 \\
	1 & 0.240 & 0.938 & 1.770 & 2.389 & 0.836 & 3.448 \\
	2 & 0.195 & 0.835 & 1.467 & 2.430 & 0.937 & 4.897 \\
	3 & 0.152 & 0.737 & 1.299 & 2.470 & 1.056 & 6.986 \\
	\hline
 \end{tabular}
 \caption{Fitting parameters $\alpha$, $s$, $s'$, $v_0$ for binary velocity distribution 
(eq.~\ref{eq:flogv_fitting_function}) for different absolute magnitude limits 
$M_V$. We also tabulate the normalization $\mathcal{N}$ and binary dispersion 
$\sigma_b$ assuming the fiducial period distribution parameters $\mu_{\log P} = 
2.23$, $\sigma_{\log P} = 2.3$ ($P$ in years).}
\label{fit_parameter_table}
\end{table}

The fitting parameters are tabulated in table \ref{sec:fit_parameter_table} for 
different absolute magnitude limits $M_{lim}$; for magnitude limits in between 
the tabulated values, interpolation may be used to find the fitting parameters.  
We also tabulate the normalization $\mathcal{N}$ and binary dispersion 
$\sigma_b$ assuming the fiducial period distribution parameters $\mu_{\log 
P}=2.23$, $\sigma_{\log P}=2.3$ ($P$ in years). While these parameter values 
were chosen with $v$ in units of AU/year, converting to other units simply 
translates $f_b(\log|v|)$ along the $\log|v|$-axis by an amount $\Delta 
\log|v|= \log \kappa$, where $\kappa$ is the conversion factor from AU/year.  
The conversion factor to km/s is $\kappa = 4.741$.

The function $f_b(\log|v|)$ in eq.~\ref{eq:flogv_fitting_function} can be 
conveniently used in place of eq.~\ref{eq:fb_logv}, which is computationally 
difficult.  Likewise, it can be substituted in eq.~\ref{fv_binary_part} and 
integrated numerically to calculate the likelihood of binary stars in a galaxy 
of given intrinsic dispersion $\sigma$. 

If the period distribution parameters $\mu_{\log P}$ and $\sigma_{\log P}$ are 
varied from their fiducial values, the fitting parameters remain unchanged to 
first order (although the normalizing factor changes according to eqs.
~\ref{normalizing_factor} - \ref{normalizing_x_factor}). Thus we have the 
additional benefit that the binary dispersion can be calculated and its 
dependence on the period distribution parameters is seen explicitly.  Analytic 
formulas for the binary dispersion are given in appendix \ref{sec:appendix_c}.  
Here we shall simplify matters by taking a log-normal distribution so that 
$s=s'$; in this case, we find the approximation dispersion is proportional to

\begin{equation}
\sigma_b^2 \propto \exp\left[2\left(\frac{\sigma_{\log P}\ln 
10}{3s}\right)^2\right] \exp\left[-2\frac{(\mu_{\log P} + 3\alpha)\ln 
10}{3s}\right]
\end{equation}

Thus the variation of $\sigma_b$ with the period parameters is quite distinct 
from its variation with $B$, which is linear (eq.~\ref{dispersion_total}). If 
one finds the dispersion up to a cutoff velocity (of order AU/year), however, 
extra factors are introduced (see eq.~\ref{sigma_b_cutoff_approx} in appendix 
\ref{sec:appendix_c}) which mitigate the exponential dependence with 
$\sigma_{\log P}$ and $\mu_{\log P}$ and nearly linearize them over their 
viable range of values (roughly from 0-4 with $P$ in years; see section 
\ref{sec:period_dist}).  Thus if one chooses a cutoff $v_c = c\sigma$ where 
$\sigma$ is the measured dispersion (e.g. $c=3$), the dispersion is 
approximately linear in the period parameters, consistent with the expected 
near-degeneracy with binary fraction.  This reinforces the very useful 
observation that the effect of binaries on the measured dispersion of a galaxy 
can be discerned from multi-epoch data in a model-independent way.

\section{Constraining the distribution of periods}\label{sec:period_dist}

Here we address the question of what form the distribution of periods might 
take in regions beyond the solar neighborhood, and whether this can be 
constrained by radial velocity data.  Simulations of star formation 
(\citealt{machida2009}, \citealt{tohline2002}, \citealt{bate2000}) indicate 
that the statistical properties of binary systems are determined during star 
formation via turbulent fragmentation of a rotating gas cloud. The 
distributions in the orbital parameters generally undergo little subsequent 
modification by collisional processes, except in the high-density regions found 
at the centers of globular clusters (\citealt{hut1992}, \citealt{pryor1988}).  
This suggests that the distribution of periods in dwarf spheroidals and dwarf 
irregular galaxies may be of a similar form to that found in the solar 
neighborhood.  We shall assume the log-normal is a fair approximation to the 
period distribution; it is reasonable to ask, however, to what extent its mean 
$\mu_{\log P}$ and dispersion $\sigma_{\log P}$ may be expected to differ from 
that of the solar neighborhood.  Because of the difficulty of simulating binary 
star formation, at present we have an incomplete picture of how these 
parameters might vary depending on the star formation history of each galaxy.  
However, the following points can be made.

A semi-empirical model of isolated binary star formation by \cite{fisher2004} 
yielded values of $\sigma_{\log P}$ within the range 1.6-2.1, depending on the 
star formation efficiency of the initial gas cloud.  Observations of pre-main 
sequence stars in Milky Way stellar associations also show more peaked 
distributions than $\sigma_{\log P} = 2.3$ (\citealt{brandner1998}).  This 
suggests that the wider distribution observed in the solar neighborhood may be 
formed by a superposition of more sharply peaked binary distributions resulting 
from various star-forming environments. Accordingly, to be conservative in this 
paper we have considered values of $\sigma_{\log P}$ $\in (0.5,4)$.  Likewise 
we have considered values of $\mu_{\log P} \in (-1,4)$ ($P$ in years).  The 
low-$\mu_{\log P}$, high-$\sigma_{\log P}$ boundaries of these intervals 
describe velocity distributions with a highly distorted, non-Gaussian shape and 
therefore can be considered unlikely.

\begin{figure}
	\includegraphics[height=1.0\hsize,angle=-90]{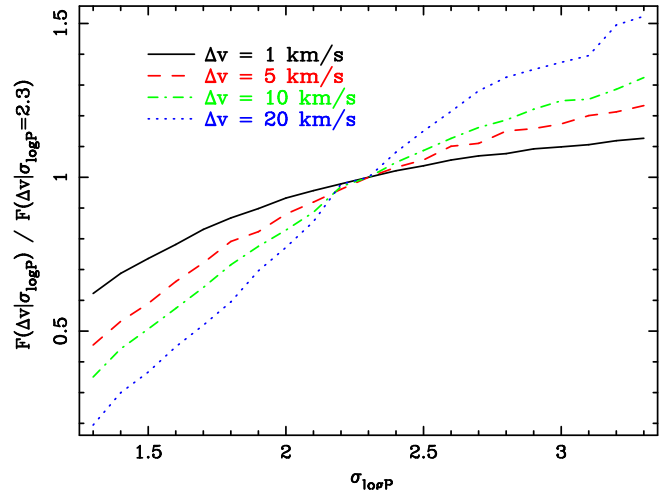}
\caption{The behavior of the threshold fraction $F(\Delta v)$ if the width of 
the period distribution $\sigma_{\log P}$ (P in years) is varied, plotted for 
different velocity thresholds $\Delta v$. The $y$-axis is given by $f = F/F_0$, 
where $F_0$ is the fiducial threshold fraction assuming the value $\sigma_{\log 
P} = 2.3$ observed in the solar neighborhood (\citealt{duquennoy1991}). }
\label{fig:Fdv_sig}
\end{figure}

\begin{figure}
	\includegraphics[height=1.0\hsize,angle=-90]{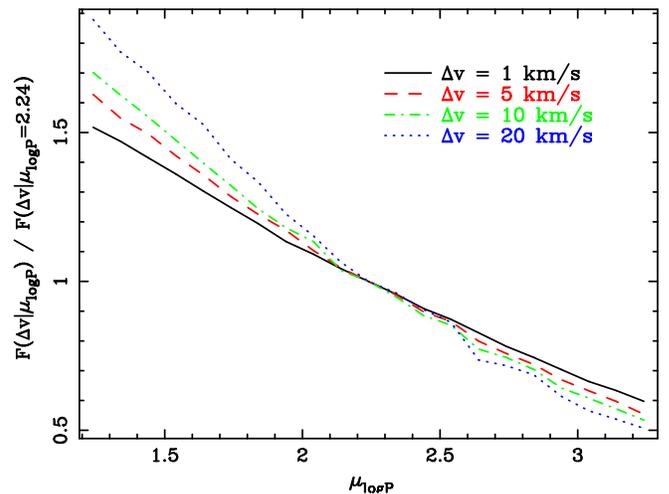}
\caption{The behavior of the threshold fraction $F(\Delta v)$ if the mean of 
the period distribution $\mu_{\log P}$ (P in years) is varied, plotted for 
different velocity thresholds $\Delta v$. The $y$-axis is given by $f = F/F_0$, 
where $F_0$ is the fiducial threshold fraction assuming the value $\mu_{\log P} 
= 2.23$ observed in the solar neighborhood (\citealt{duquennoy1991}). }
\label{fig:Fdv_mu}
\end{figure}

Supposing that $\mu_{\log P}$ and $\sigma_{\log P}$ may vary by the amount 
suggested by \cite{fisher2004}, can these parameters be estimated independently 
of binary fraction? To answer this question, we reconsider the threshold 
fraction $F(\Delta v|\Delta t)$, defined as the fraction of stars with an 
observed change in velocity greater than a threshold $\Delta v$ after a time 
$\Delta t$ between measurements.  For each threshold, let us define the ratio 
$f = F/F_0$ where $F_0$ is the fiducial threshold fraction obtained by setting 
$\sigma_{\log P}=2.3,\mu_{\log P}=2.23$.  Given a time interval of 1 year and 
picking several thresholds $\Delta v$, we plot $f$ as a function of 
$\sigma_{\log P}$ in fig.~\ref{fig:Fdv_sig} and $\mu_{\log P}$ in 
fig.~\ref{fig:Fdv_mu}. The similarity in slope among the different curves give 
a measure of degeneracy; if the curves were identical, the parameters would be 
completely degenerate with the binary fraction.  However, if the difference 
between the curves is smaller than the statistical error bars, the parameters 
are effectively degenerate with $B$ and we cannot hope to distinguish between 
them.  The error bars on $f$ at a particular $\Delta v$ are given by (compare 
eq.~\ref{sigF_best_fit} in appendix \ref{sec:appendix_b}):

\begin{equation}
\epsilon_f \approx \frac{2 f}{\sqrt{N \bar F(\Delta v_{tail}|\Delta t)}}
\end{equation}
where $\epsilon_f$ is the 95\% confidence limit in $f$, $N$ is the number of 
stars and $\Delta v_{tail}$ is defined as before (for a rough approximation, 
given a measurement error $\sigma_m$ one can take $\Delta v_{tail} \approx 
2\sigma_{2e}$ where $\sigma_{2e} \approx \sqrt{2}\sigma_m$.  For $\Delta t$ = 1 
year and a measurement error of 2 km/s, fig.~\ref{fig:dvchist_dt} gives $\bar 
F(\Delta v_{tail}) \approx 0.1\times B$.  If $B \approx 0.5$, this gives 
$\epsilon_f \approx 9f/\sqrt{N}$.  Assuming a sample of $N=100$ stars, we have 
$\epsilon_f \approx 0.9$, far too large to distinguish between the different 
curves in figs.~\ref{fig:Fdv_sig} and \ref{fig:Fdv_mu}. Thus to within the 
statistical error in a sample of a few hundred stars, $\sigma_{\log P}$ and 
$\mu_{\log P}$ are degenerate with the binary fraction over nearly the entire 
range of figs.~\ref{fig:Fdv_sig} and \ref{fig:Fdv_mu}; only with $N \gtrsim 
1000$ are the error bars small enough to break this degeneracy. We conclude 
that multi-epoch samples of $\approx 1000$ stars or more are required in order 
to constrain the period distribution of a population of binary systems.

%\begin{figure}
	%\includegraphics[height=1.0\hsize,angle=-90]{deltav.cumulative_dist.ps}
%\caption{Cumulative distributions of $\Delta v$ after a 1-year interval, for 
%binary fractions of 1.0 (green) and 0.5 (red). Dashed line is the cumulative 
%distribution for 175 stars in Fornax taken from the Magellan survey 
%(\cite{walker2009}), whose radial velocities were measured roughly 1 year 
%apart. Solid curves have a measurement error of 1.4 km/s for all data points 
%(the robust average error in the Magellan sample is 1.36 km/s), while 
%short-dotted lines have zero measurement error.
%}
%\label{fig:cumulative_dist}
%\end{figure}

\section{Conclusion}\label{sec:conclusion}

We have demonstrated a procedure to estimate the intrinsic velocity dispersion 
of dwarf spheroidal galaxies more precisely than in previous studies by 
accounting for the effect of binary orbital motion.  By measuring the 
\emph{threshold fraction} of a stellar sample (defined in section 
\ref{sec:threshold_fraction}), the correction to the velocity dispersion due to 
binary motion can be estimated; fitting functions are provided for this method 
(eqs.~\ref{eq:sigratio_fitting_function}, 
\ref{eq:sigratio_fit_parameter_a}-\ref{eq:sigratio_fit_parameter_c}). We have 
also demonstrated a method to estimate the threshold fraction, together with 
other properties of the binary population, more rigorously from multi-epoch 
data via a Bayesian or maximum likelihood approach.  We conclude with the 
following points:

1. The dispersions of dwarf spheroidal galaxies between 4-10 km/s are unlikely 
to be inflated by more than 20\% by binaries (fig.~\ref{fig:fdisp_sigreal}).  
This conclusion can be made with confidence because---as we showed in section 
\ref{sec:correcting_dispersion}---the correction to the dispersion holds 
independently of the model used to describe the binary population, provided 
that the dispersion is calculated using a high-velocity cutoff (e.g., by a 
3$\sigma$-clipping routine).  This is fortunate since, as we showed in section 
\ref{sec:period_dist}, the binary fraction and other properties of the binary 
population (e.g., period distribution) are very difficult to constrain 
independently of each other with samples of only a few hundred stars.

2. With a measurement error on the order of 1 km/s, we find that 1-2 years is 
an optimal interval between measurements for measuring the threshold fraction, 
since the fraction of stars with a measureable change in velocity does not rise 
significantly when the interval is extended beyond 2 years 
(fig.~\ref{fig:dvchist_dt}).  Furthermore, to constrain the binary fraction and 
other properties of the binary population, it is a more profitable strategy to 
make two-epoch measurements over a larger sample of stars, as opposed to adding 
more repeat measurements to an already existing two-epoch sample. This is 
necessary to overcome the large scatter in the binary fraction of samples with 
less than a few hundred stars.  We also find that multi-epoch samples of more 
than 1000 stars would be required in order to constrain the distribution of 
periods of a binary population independently of the binary fraction.

3. While the binary population can in principle be constrained by single-epoch 
data, in practice this is very difficult because of contamination by non-member 
stars.  Since an outlier in the velocity distribution cannot be verified as a 
binary star without multiple epoch measurements, it may be erroneously labeled 
a non-member star and excluded from the data sample (or weighted by a low 
membership probability). This would bias the estimated binary fraction to low 
values, resulting in an inflated dispersion estimate.  Even in single-epoch 
samples, however, this problem may be overcome by including a likelihood for 
the foreground Milky Way stars in a Bayesian analysis. This analysis can also 
be combined with multi-epoch data to provide better constraints, and in 
principle would extract the best constraints from both single- and multi-epoch 
velocity data.

\section*{Acknowledgements}
We would like to thank Erik Tollerud for providing valuable feedback and many 
illuminating discussions throughout the course of this project. This work was 
supported in part by NSF grant AST-0607746 and NASA grant NNX09AD09G.

\bibliographystyle{apj}
\bibliography{binary}

\setcounter{section}{0}
\section{Appendix: Calculating the threshold fraction for different measurement 
errors}\label{sec:appendix_a}

In section \ref{sec:correcting_dispersion} we showed how to correct the dispersion 
(eq.~\ref{eq:sigratio_fitting_function}) by using the threshold fraction $F$, 
defined as the fraction of stars with an observed change in velocity greater 
than a threshold $\Delta v$ after a time $\Delta t$ between measurements. If 
the threshold fraction is calculated directly from the data, then measurement 
error must be considered in eq.~\ref{eq:sigratio_fitting_function}.  While the 
measured dispersion can be easily corrected for measurement error according to 
$\sigma_{meas}^2 = \sigma^2 - \sigma_m^2$ (where $\sigma_{meas}$ is the 
measured dispersion and $\sigma_m$ is the measurement error), correcting the 
threshold fraction is less trivial. To correct the threshold fraction for 
measurement error, first we note that the degeneracy of the period parameters 
with binary fraction means that the velocity distribution can be approximately 
split into two parts: a small-$v$ part which acts effectively like a 
$\delta$-function similar to that in eq.~\ref{eq:f_dv_noerr}, and a large-$v$ 
tail which scales linearly with $B$, $\mu_{\log P}$, and $\sigma_{\log P}$.  
Exactly where to do the ``splitting'' is dictated largely by the measurement 
error, as the tail becomes prominent only at velocities beyond $\Delta v 
\approx \sigma_{2e}$. We therefore split the function at $\Delta v_{tail} = 
\gamma \sigma_{2e}$ where $\gamma$ is a proportionality constant with a very 
weak (if any) dependence on $\sigma_{2e}$.  The threshold fraction can then be 
written approximately as follows:

\begin{equation}
\label{eq:F_dv_approx}
F(\Delta v|\Delta t,B,\sigma_{2e}) \approx (1 - 
\mathcal{N})\textrm{erfc}\left[\frac{\Delta v}{\sqrt{2}\sigma_{2e}}\right] + 
BJ(\Delta v|\sigma_{2e})
\end{equation}
where
\begin{equation}
\label{eq:J_function}
J(\Delta v|\sigma_{2e}) = 
\int_{-\infty}^{\infty}\textrm{erfc}\left[\frac{\Delta v-\Delta 
v'}{\sqrt{2}\sigma_{2e}}\right] \frac{g_b(\log|\Delta v'|)}{|\Delta v'|\ln 
10}d(\Delta v'),
\end{equation}

\begin{equation}
\mathcal{N} = B J(\gamma \sigma_{2e}|\sigma_{2e}).
\end{equation}

The normalization factor $\mathcal{N}$ here has replaced $B$ in 
eq.~\ref{eq:F_dv} and varies linearly with $B$, $\mu_{\log P}$ and 
$\sigma_{\log P}$. Now as long as $\Delta v$ is approximately equal to or 
larger than $\sigma_{2e}$, the error function in the integrand of 
eq.~\ref{eq:J_function} is approximately a smoothed step function. For $\Delta v 
/ \sigma_{2e} \gtrsim 1$ we can therefore make the approximation

\begin{equation}
\label{eq:J_function_approx}
J(\Delta v|\sigma_{2e}) \approx \alpha\left(\frac{\Delta v}{\sigma_{2e}}\right) F_0(\Delta v)
\end{equation}
where $F_0(\Delta v) \equiv F(\Delta v|\sigma_{2e}=0)$ denotes the threshold 
fraction without measurement error.  Note that in the limit $\sigma_{2e} 
\rightarrow 0$ ~ ($\Delta v/\sigma_{2e} \rightarrow \infty$), the integrand of 
eq.~\ref{eq:J_function} becomes a step function so that $\alpha \rightarrow 1$.  
On the other end, as $\Delta v/\sigma_{2e}$ is made smaller, the integral in 
eq.~\ref{eq:J_function} includes more of the central peak so $\alpha$ becomes 
larger. Using eq.~\ref{eq:J_function_approx}, we can write $\mathcal{N} = 
\kappa F_0(\Delta v)$ where

\begin{equation}
\kappa = \frac{\alpha(\gamma) F_0(\gamma\sigma_{2e})}{\alpha\left(\frac{\Delta v}{\sigma_{2e}}\right) F_0(\Delta v)}
\end{equation}

Since $F_0$ is the tail of a log-normal distribution, over the scale of several 
km/s we have $F_0(\Delta v) \approx (\Delta v)^{-n}$ where $n$ is an exponent 
close to 1. Therefore $\kappa$ can be written as a function of $\Delta v / 
\sigma_{2e}$:

\begin{equation}
\kappa\left(\frac{\Delta v}{\sigma_{2e}}\right) ~ \approx ~ \frac{\gamma^{-n}\alpha(\gamma)}{\alpha\left(\frac{\Delta v}{\sigma_{2e}}\right)} \left(\frac{\Delta v}{\sigma_{2e}}\right)^n
\end{equation}

Substituting this result into eq.~\ref{eq:F_dv_approx}, we have

\begin{equation}
\label{eq:F_measurement_error_key_eq_appendix}
F(\Delta v|\sigma_{2e}) \approx \textrm{erfc}\left[\frac{\Delta 
v}{\sqrt{2}\sigma_{2e}}\right] + \beta\left(\frac{\Delta v}{\sigma_{2e}}\right) 
F_0(\Delta v)
\end{equation}
where
\begin{equation}
\beta\left(\frac{\Delta v}{\sigma_{2e}}\right) \equiv \alpha\left(\frac{\Delta 
v}{\sigma_{2e}}\right) \left\{1 - \kappa\left(\frac{\Delta 
v}{\sigma_{2e}}\right)\textrm{erfc}\left[\frac{\Delta 
v}{\sqrt{2}\sigma_{2e}}\right]\right\}.
\end{equation}

Eq.~\ref{eq:F_measurement_error_key_eq_appendix} is the key result: it means 
that the threshold fraction $F$ with a given measurement error is related to 
the measurement error-free value $F_0$ by a linear transformation, and the only 
extra information required to make this transformation is the function 
$\beta\left(\frac{\Delta v}{\sigma_{2e}}\right)$. As a check, taking the limit 
$\sigma_{2e} \rightarrow 0$ we have $\alpha \rightarrow 1$ and $\kappa 
\rightarrow 0$, so that $\beta \rightarrow 1$ as it should. We use the Monte 
Carlo simulation to plot $\beta$ as a function of $\Delta v/\sigma_{2e}$ for 
different velocity thresholds, and find the function $\beta$ is nearly the same 
regardless of threshold, which justifies the approximations taken to reach 
eq.~\ref{eq:F_measurement_error_key_eq}. We find that for $\Delta v/\sigma_{2e} 
\gtrsim 1$, the parameter $\kappa$ shows only slight variation over the range 
for which the error function is non-negligible. We also find that $\alpha$ is 
well approximated by an exponential, so that $\beta$ is well fit by the 
following function:

\begin{equation}
\beta(x) = \left(1 + a e^{-\frac{x}{x_s}}\right) \left\{1 - \bar \kappa \cdot
\textrm{erfc}\left(\frac{x}{\sqrt{2}}\right)\right\}
\end{equation}
where $x = \Delta v/\sigma_{2e}$ and the best-fit parameters are $a = 0.05$, 
$x_s = 5$, and $\bar \kappa = 1.3$.

\section{Appendix: Formula for number of stars required to constrain the binary 
fraction}\label{sec:appendix_b}

For a set of $N$ stellar velocities (with $N > 100$) measured at two different 
epochs separated by time $\Delta t$, consider the number of stars $n$ with 
change in velocity greater than some threshold value $\Delta v$. If $n$ were to 
be measured over many random realizations drawn from a particular galaxy, it 
would follow a Poisson distribution. For the time being, let us take the 
measurement error $\sigma_{2e}$ to be zero.  The mean number $\bar n$ is 
related to the mean threshold fraction of binaries $\bar F_b(\Delta v|\Delta 
t,\sigma_{2e}=0)$ by $\bar n = N B \bar F_b$ where $N$ is the total number of 
stars. If we pick $\Delta v$ small enough such that this number is larger than 
10, the Poisson distribution is approximately Gaussian with standard deviation 
$\sigma_n \approx \bar n \approx \sqrt{N B \bar F_b}$. Since the fraction of 
stars is $F = n/N$, we obtain the error in the measured threshold fraction $F$:

\begin{equation}
\sigma_F = \sqrt{\frac{\bar F(\Delta v|\Delta t,B,\sigma_{2e}=0)}{N}}
\label{sigma_F_eq}
\end{equation}
where we also used the relation $\bar F = B \bar F_b$. Now if $\sigma_{2e}$ is 
nonzero, the smallest value of $\Delta v$ which is largely unaffected by the 
measurement error will occur roughly at the point $\Delta v_{tail}$ where $\bar 
F(\Delta v_{tail}|\Delta t,B,\sigma_{2e}=0) = \bar F(\Delta v_{tail}|\Delta 
t,B=0,\sigma_{2e})$; this is where the ``tail'' in the distribution due to 
binaries begins to dominate. Therefore we pick this point as giving the best 
constraint on $B$. (For a rough approximation, one can also use $\Delta 
v_{tail} \approx 2\sigma_{2e}$.) If we measure the fraction of stars with 
change in velocity greater than $\Delta v_{tail}$, given by $F = n/N$, our 
``best-fit'' binary fraction $b$ is then defined by $F = b \bar F_b(\Delta 
v_{tail}|\Delta t,B,\sigma_{2e}=0)$.  Combining this with $\bar F = B \bar F_b$ 
and eq.~\ref{sigma_F_eq} gives the standard deviation of the best-fit binary 
fraction:

\begin{equation}
\sigma_b = \sqrt{\frac{B}{N \times \bar F_b(\Delta v_{tail}|\Delta t,\sigma_{2e}=0)}}
\label{sigB_N}
\end{equation}

Now calling the 95\% confidence limit $\epsilon_b = 2\sigma_b$ and solving for 
$N$ yields eq.~\ref{eq:N_sigB}. The approximation becomes less accurate as $B$ 
tends towards very small (close to zero) and large (close to one) values. If 
$B$ is sufficiently close to zero such that $n$ is less than 10, the Gaussian 
limit no longer holds; in that limit, the uncertainty will be larger than that 
given in eq.~\ref{sigB_N}. If $B$ is close to 1, the width of the Gaussian is 
larger than the true uncertainty since $B$-values greater than 1 are not 
allowed. In fig.~\ref{fig:approx} the approximation formula is graphed as a 
function of $N$ for several values of $B$ and compared to the 95\% confidence 
interval of the posterior $P(B)$ obtained from a Bayesian analysis of the 
simulated data, averaged over a hundred realizations.

Using the best-fit binary fraction obtained by the threshold fraction at 
$\Delta v_{tail}$, we can find the standard deviation of the best-fit threshold 
fraction at a given threshold $\Delta v$ by substituting the relations $F = b 
\bar F_b(\Delta v|\Delta t,B,\sigma_{2e}=0)$ and $\bar F = B \bar F_b$ into 
eq.~\ref{sigB_N}, with the result

\begin{equation}
\label{sigF_best_fit}
\sigma_{F,fit} = \frac{\bar F(\Delta v|\Delta t, \sigma_{2e}=0)}{\sqrt{N\bar 
F(\Delta v_{tail}|\Delta t,\sigma_{2e}=0)}}
\end{equation}

This equation differs from eq.~\ref{sigma_F_eq} in that it uses the best-fit 
binary fraction to infer the threshold fraction at thresholds $\Delta v > 
\Delta v_{tail}$. Since the scatter in $F$ is smaller at $\Delta v_{tail}$, 
this leads to a better constraint than if the threshold fraction is measured 
directly. Comparing eqs.~\ref{sigF_best_fit} and \ref{sigma_F_eq} and using the 
fact that the threshold fraction $F$ decreases monotonically in $\Delta v$, we 
see that the error $\sigma_{F,fit} < \sigma_{F}$ as expected.  
Eq.~\ref{sigF_best_fit} approximates the error in the threshold fraction 
estimated by the Bayesian approach outlined in section \ref{sec:multi_epoch}.  
%We caution, however, that if one uses high thresholds ($\Delta v > 10$ km/s), 
%the inferr

\section{Appendix: Analytic formulas for binary velocity dispersion} 
\label{sec:appendix_c}

For the purpose of making analytic calculations or doing a likelihood analysis, 
it is useful to have a fitting function for the binary velocity distribution; 
this function was defined in section \ref{sec:fitting_function} 
(eqs.~\ref{eq:flogv_fitting_function}-\ref{eq:g_fitting_function}). If the 
period distribution parameters $\mu_{\log P}$ and $\sigma_{\log P}$ and are 
varied from their fiducial values, the fitting parameters remain unchanged to 
first order (although the normalizing factor changes according to eqs.
~\ref{normalizing_factor} - \ref{normalizing_x_factor}).  Thus we have the 
additional benefit that the binary dispersion can be calculated and its 
dependence on the period distribution parameters is seen explicitly. In terms 
of the fitting parameters given in 
eqs.~\ref{eq:flogv_fitting_function}-\ref{eq:g_fitting_function} and tabulated 
in table \ref{fit_parameter_table}, we find the following analytic formula for 
the binary dispersion:

\begin{equation}
\sigma_b^2 = \frac{\mathcal{N}\kappa^2}{2}\left\{f(s)\textrm{erfc}[-x(v_0;s)] + f(s')\textrm{erfc}[x(v_0;s')]\right\}
\end{equation}
where $\kappa = 4.741$ km/s is the conversion factor from AU/year, and

\begin{equation}
f(s) = \frac{1}{s}\exp\left[2\left(\frac{\sigma_{\log P}\ln 10}{3s}\right)^2\right]\exp\left[\frac{-2(\mu_{\log P}+3\alpha)\ln 10}{3s}\right],
\label{f_s}
\end{equation}

\begin{equation}
x(v;s) = \frac{1}{\sqrt{2}\sigma_{\log P}}\left[3s\log|v| + \mu_{\log P} + 3\alpha - \frac{2}{3s}\sigma_{\log P}^2\ln 10\right].
\label{x_s}
\end{equation}

Although the normalization $\mathcal{N}$ also varies with the period 
distribution parameters $\mu_{\log P}$ and $\sigma_{\log P}$ according to 
eqs.~\ref{normalizing_factor}-\ref{normalizing_x_factor}, this variation is 
weak compared to the exponential dependence in $f(s)$ (eq.~\ref{f_s}). The 
binary dispersion $\sigma_b$ is tabulated in table \ref{fit_parameter_table} 
assuming the fiducial period distribution parameters $\mu_{\log P}=2.23$, 
$\sigma_{\log P}=2.3$ ($P$ in years).

One can also find the dispersion with a velocity cutoff $v_c$, according to 
which stars with $v > v_c$ are discarded from the sample. We denote the 
dispersion with a cutoff as $\sigma(v_c)$. A little algebra shows that

\begin{equation}
\sigma^2(v_c) = \sigma_0^2(v_c) + B \sigma_b^2(v_c)
\end{equation}
where
\begin{equation}
\label{eq:sigma_0_cutoff}
\sigma_0^2(v_c) = \sigma_0^2\textrm{erf}\left[\frac{v_c}{\sigma_0\sqrt{2}}\right] - \sigma_0 v_c \sqrt{\frac{2}{\pi}} e^{-v_c^2/2\sigma_0^2},
\end{equation}

\begin{equation}
\label{eq:sigma_b_cutoff}
\sigma_b^2(v_c) = \frac{1}{2}\int_{-\infty}^\infty \left\{\textrm{erfc}\left[\frac{v_c-|v'|}{\sigma_0\sqrt{2}}\right] + \textrm{erfc}\left[\frac{v_c+|v'|}{\sigma_0\sqrt{2}}\right]\right\}|v'|^2 f_b(\log|v'|)d\log|v'|.
\end{equation}

If a cutoff is chosen such that $v_c \gtrsim 2.7\sigma_0$, then the first term 
dominates in eq.~\ref{eq:sigma_0_cutoff}, and the first term in the integrand 
of eq.~\ref{eq:sigma_b_cutoff} dominates over the second term. The 
complementary error function can be approximated as a step function, in which 
case the integral can be done analytically with the result

\begin{equation}
\sigma_b^2(v_c) \approx \frac{\mathcal{N}\kappa^2}{2}\left\{f(s)\left[1 + \textrm{erf}|x(v_0;s)|\right] + f(s')\left[\textrm{erf}|x(v_c;s')| - \textrm{erf}|x(v_0;s')|\right]\right\}
\label{sigma_b_cutoff_approx}
\end{equation}

The extra factor in eq.~\ref{sigma_b_cutoff_approx} introduced by the cutoff 
$v_c$ mitigates the exponential dependence of $f(s)$ with the period 
distribution parameters $\sigma_{\log P}$ and $\mu_{\log P}$; indeed it nearly 
linearizes them over their viable range of values (roughly from 0-4 with $P$ in 
years; see section \ref{sec:period_dist}).  Thus if one chooses a cutoff $v_c = 
c\sigma$ where $\sigma$ is the measured dispersion (e.g.  $c=2.7$), the 
dispersion is approximately linear in the period parameters, consistent with 
the expected near-degeneracy with binary fraction.  This reinforces the very 
useful observation that the effect of binaries on the measured dispersion of a 
galaxy can be discerned from multi-epoch data in a model-independent way.

\end{document}